\documentclass[a4paper,fleqn, 11pt]{cas-sc} 

\usepackage{graphicx}
\usepackage{wrapfig}
\usepackage{amssymb}
\usepackage{amsmath}
\usepackage{xcolor}
\usepackage{amsthm}
\usepackage{amsmath}
\usepackage{subcaption}
\usepackage{balance}
\usepackage{hyperref}
\usepackage[export]{adjustbox}
\def\tsc#1{\csdef{#1}{\textsc{\lowercase{#1}}\xspace}}
\tsc{AU}
\tsc{FJDG}

\definecolor{editor}{RGB}{0,128,128}

\definecolor{softred}{RGB}{200, 50, 50}
\colorlet{softblue}{blue!90!black}
\usepackage[square,numbers,sort,compress]{natbib}
\usepackage{nameref}
\begin{document}

\shorttitle{ Ductile-to-Brittle Transition}

\shortauthors{A.\ Ustrzycka et~al.}

\title [mode = title]{\textcolor{black}{Atomistic Study of Radiation-Induced Ductile-to-Brittle Transition in Austenitic Steel}}
                   



\author[1]{A. Ustrzycka}[type=editor,
                        orcid=]
\cormark[1] 

\author[1]{H. Mousavi}[type=editor,
                        orcid=]

\author[2]{F. J. Dominguez-Gutierrez}
\credit{}

\author[1]{S. Stupkiewicz}
\credit{}

\affiliation[1]{organization={Institute of Fundamental Technological Research, Polish Academy of Sciences},
    addressline={Pawinskiego 5B}, 
    city={Warsaw},
    postcode={02-106}, 
    country={Poland}}

\affiliation[2]{organization={NOMATEN Centre of Excellence, National Centre for 
Nuclear Research},
    addressline={ul. A. Soltana 7}, 
    city={Otwock},
    postcode={05-400}, 
    country={Poland}}

\begin{abstract} 
Neutron irradiation in structural alloys promotes defect clustering, which suppresses plasticity and triggers a ductile-to-brittle transition (DBT), a key degradation mechanism limiting fracture resistance in nuclear materials. This study investigates the fracture mechanisms underlying this transition in irradiated Fe--Ni--Cr alloys. Using Molecular Dynamics simulations, we examine how different defect types influence crack propagation and energy dissipation mechanisms. The results reveal distinct roles of these defects: voids facilitate crack growth by reducing local cohesive energy, while dislocation loops act as barriers that impede crack advancement and redirect crack paths, significantly altering crack morphology.
Building on the classical approach of separating fracture energy into brittle cleavage and plastic components, this study adapts the decomposition to irradiated materials. This framework quantifies the evolving contributions of surface energy and plastic work across increasing radiation damage levels, providing critical insight into how irradiation-induced defects govern fracture dynamics.
\end{abstract}




\maketitle
\section{Introduction}
\label{sec: Section2}
\vspace{0.3cm}
The embrittlement of metals, resulting from reduced plastic deformation preceding fracture, is a significant determinant of the durability and safety of structural materials, particularly under the extreme conditions encountered in nuclear reactors. Irradiation embrittlement plays a crucial role in the degradation of the mechanical properties of these materials \citep{Nordlund2018, LIU2023, LIN2024}. Radiation defects arise when energetic particles, such as neutrons or ions, collide with atoms in the lattice, displacing them from their original positions and creating vacancies and interstitial atoms, which subsequently cluster into more complex defect structures such as dislocations and voids \citep{Kinchin1955, NORGETT1975, PhysRevLett.116.135504}. 
\textcolor{black}{Those defects impede dislocation motion and reduce the material's capacity for plastic deformation, ultimately leading to a brittle failure. The related phenomena are collectively referred to as the radiation-induced ductile-to-brittle transition (DBT).} 

DBT is a well-known phenomenon in metals that is usually related to the temperature sensitivity of fracture stress and yield stress \citep{LI2018, BARIK2023, BARIK2024, LIN2025}. DBT is characterized by a shift from ductile fracture, which involves significant plastic deformation, to brittle fracture, where materials fail with little prior deformation \citep{CHENG2010, MOLLER2015}.
This transition is heavily influenced by temperature, stress state, and microstructural factors, particularly the presence of defects. While temperature is recognized as an important factor in the overall understanding of the DBT phenomenon, the accumulation of radiation-induced defects and their influence on mechanical properties are considered the determining elements in the case of irradiated materials. 

\begin{figure}[t!]
    \centering
    \includegraphics[scale=0.8]{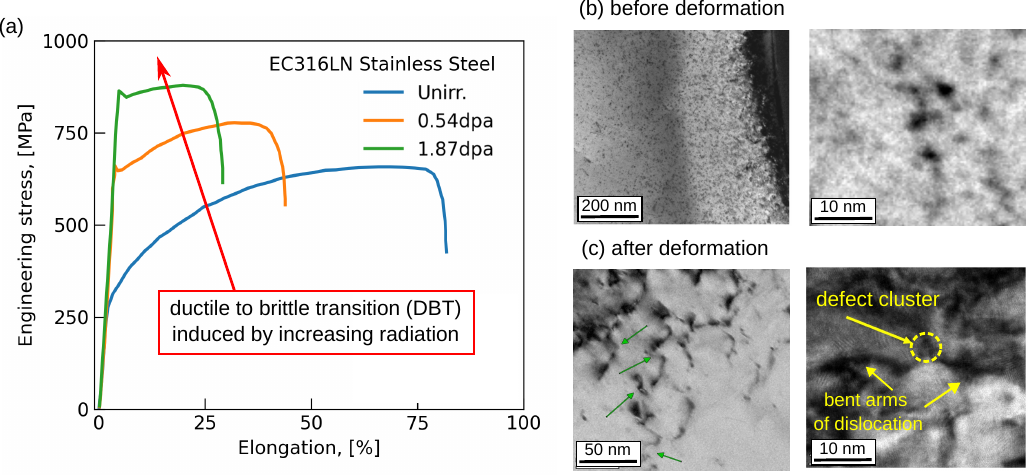}
\caption{(a) Engineering stress--strain curves of stainless steel after irradiation, demonstrating ductile-to-brittle transition (after \citet{Byun2002}).
(b) Transmission electron microscopy (TEM) analysis of steel irradiated to 0.12 dpa, showing radiation-induced porosity in the form of defect clusters at varying magnifications \citep{USTRZYCKA2024}.
(c)~TEM images illustrating the dislocation-blocking mechanism, where a dislocation line encircles radiation defects at varying magnifications \citep{USTRZYCKA2024}. These images depict the irradiated material after deformation, highlighting the interaction between dislocations and radiation-induced defects.}
    \label{fig: introduction}
\end{figure}

Fig.~\ref{fig: introduction} illustrates the impact of irradiation on the mechanical behavior of a stainless steel. The stress--strain curves demonstrate the reduction of ductility with increasing radiation-induced porosity (compare Fig.~\ref{fig: introduction}(a). The curves clearly show a decrease in elongation at fracture with higher radiation doses, indicating a significant loss of plasticity. This loss of ductility is directly related to the increase of radiation porosity and defect accumulation.
The reduction of ductility is a critical factor contributing to ductile--brittle transition (DBT). As shown in Fig.~\ref{fig: introduction}(b), irradiation-induced defects such as voids and dislocation loops form clusters that disrupt the crystalline structure, impeding the material’s ability to deform plastically \citep{Byun2001, JI2023, USTRZYCKA2024}. 
Additionally, TEM images in Fig.~\ref{fig: introduction}(c) reveal the dislocation-blocking mechanism, where dislocation lines bend and encircle radiation defects, forming curved arms. This interaction, observed after deformation, significantly restricts plastic flow by impeding dislocation motion \citep{Osetsky2004, CUI2017, BAO2022}.
These microstructural changes, driven by the accumulation of defects, significantly degrade the mechanical integrity of the material under applied stress. Thus, defect accumulation can play a pivotal role in driving DBT.

Molecular dynamics (MD) simulations, with well-known limitations concerning spatial and temporal scales, play a critical role in studying DBT by providing atomic-level insights into the mechanisms driving embrittlement \citep{ZHANG2013, XU2020, LI2024}. 
\textcolor{black}{In the case of irradiated materials, MD enables in-depth analysis of how radiation-induced defects influence crack propagation pathways and fracture behavior \citep{Krull2011, FULLER2020, JI2023, XIE2023}. It further provides insight into the mechanisms by which these defects interact with dislocations and alter their mobility \citep{WANG2015, BAO2022, TSUGAWA2022}.}
These simulations are particularly useful for examining the effects of radiation defects on material behavior, enabling the exploration of how defect accumulation changes the mechanical properties of the materials \citep{CUI2017, CHEN2020, BAO2022}. MD can simulate extreme irradiation conditions and provide a clear understanding of the transition from ductile to brittle fracture at various defect densities.
\textcolor{black}{Furthermore, MD results are essential inputs for multiscale models, enabling the development of constitutive laws that accurately represent defect-driven plasticity and fracture. That integration bridges atomic-scale mechanisms with continuum descriptions, improving predictions of damage evolution and DBT \citep{MONNET2018, ZHOU2022, KHAROUJI2024}.}

This paper presents a novel investigation into the role of radiation-induced defects, expressed by the displacement per atom (dpa) parameter \citep{Nordlund2018}, in driving DBT. Unlike previous studies, which primarily focus on temperature-driven DBT, this work examines the direct influence of irradiation-induced defects on dislocation dynamics and energy dissipation mechanisms during crack propagation. By systematically analyzing how varying dpa levels govern defect accumulation and alter the material’s capacity for plastic deformation, this study establishes a direct link between irradiation damage and fracture behavior, an aspect that has not been comprehensively addressed before. Furthermore, this paper provides the first detailed correlation between defect density and the microstructural mechanisms governing DBT. Through a~combination of MD simulations and advanced defect characterization techniques, this study offers original insights into how variations in dpa influence the mechanical response, including the transition from ductile to brittle fracture.

A comprehensive understanding of the DBT phenomenon in irradiated materials is crucial for accurately predicting material performance in nuclear applications and for devising effective strategies to mitigate embrittlement effects. Understanding the fracture behavior of irradiated structural materials is essential to ensuring the safety and reliability of structural components of nuclear reactors operating in exceptionally harsh environments due to the combination of high stresses, a chemically aggressive coolant,  high or, in the case of accelerator systems, extremely low temperatures, and intense radiation fluxes \citep{ULLMAIER1995, CHOWDHURY2016, Ustrzycka2016, ji2021atomistic}.

\textcolor{black}{The relevance of our results is particularly strong for structural materials in nuclear reactors, where neutron irradiation leads to volumetric defect accumulation throughout the bulk of the material. While the types of defects considered in this study can form under both neutron and ion irradiation, the key distinction lies in the spatial extent of damage. Our simulations are designed to reproduce the uniform, bulk-like defect distribution characteristic of neutron exposure, rather than the gradient defect profiles characteristic of layers in ion-irradiated samples.}

The paper is organized as follows: Section \ref{sec: Section2} outlines the simulation methodology and the MD framework for modeling irradiated alloys, including the preparation of irradiated samples, characterization of defect densities, and simulations of fracture behavior. 
This section focuses on the physical mechanisms underlying defect-crack interactions, including dislocation emission and crack tip shielding, to elucidate how irradiation alters fracture pathways and energy dissipation. Section \ref{sec: Section3} explores the micro-mechanisms governing DBT in irradiated materials, emphasizing the role of radiation-induced defects in altering dislocation dynamics and crack propagation. Section \ref{sec: Section4} introduces the decomposition of fracture energy into cleavage and plastic contributions to quantify the energy dissipation during fracture in irradiated materials.

\section{Simulation methodology}
\label{sec: Section2}
\textcolor{black}{This section outlines the molecular dynamics simulation approach used to investigate the effects of irradiation-induced defects on fracture behavior in Fe--Ni--Cr alloys. It is divided into two parts: first, the preparation of irradiated samples with controlled displacement damage levels is described; second, the methodology for simulating crack propagation in these pre-damaged samples is presented.} 
\subsection{Preparation of the irradiated samples}
\label{sec: Section2.1}
\vspace{0.3cm}
The methodology employed for samples irradiation closely follows procedures established in our earlier papers \citep{Dominguez2020, Toussaint2021, Dominguez2022, USTRZYCKA2024} and is only briefly summarized here. To investigate the formation and evolution of dislocations and voids under various dpa values, MD simulations were performed using the Large-scale Atomic/Molecular Massively Parallel Simulator (LAMMPS) software \citep{LAMMPS}.

The interatomic potential used for these simulations was developed by \citet{Bonny2011, Bonny_2013} based on the Embedded Atom Method (EAM) with Ziegler-Biersack-Littmark (ZBL) corrections. This potential is well-suited for modeling defect formation and short-range interactions in face-centered cubic (fcc) materials, offering reliable predictions of dislocation behavior and stacking fault energies (SFEs) over a broad temperature range of 0--900 K for Fe--Ni--Cr alloy structures \citep{BELAND2017, Dominguez2022, ZHOU2023, USTRZYCKA2024}.

\begin{figure}[t!]
\centering
\includegraphics[scale=0.5]{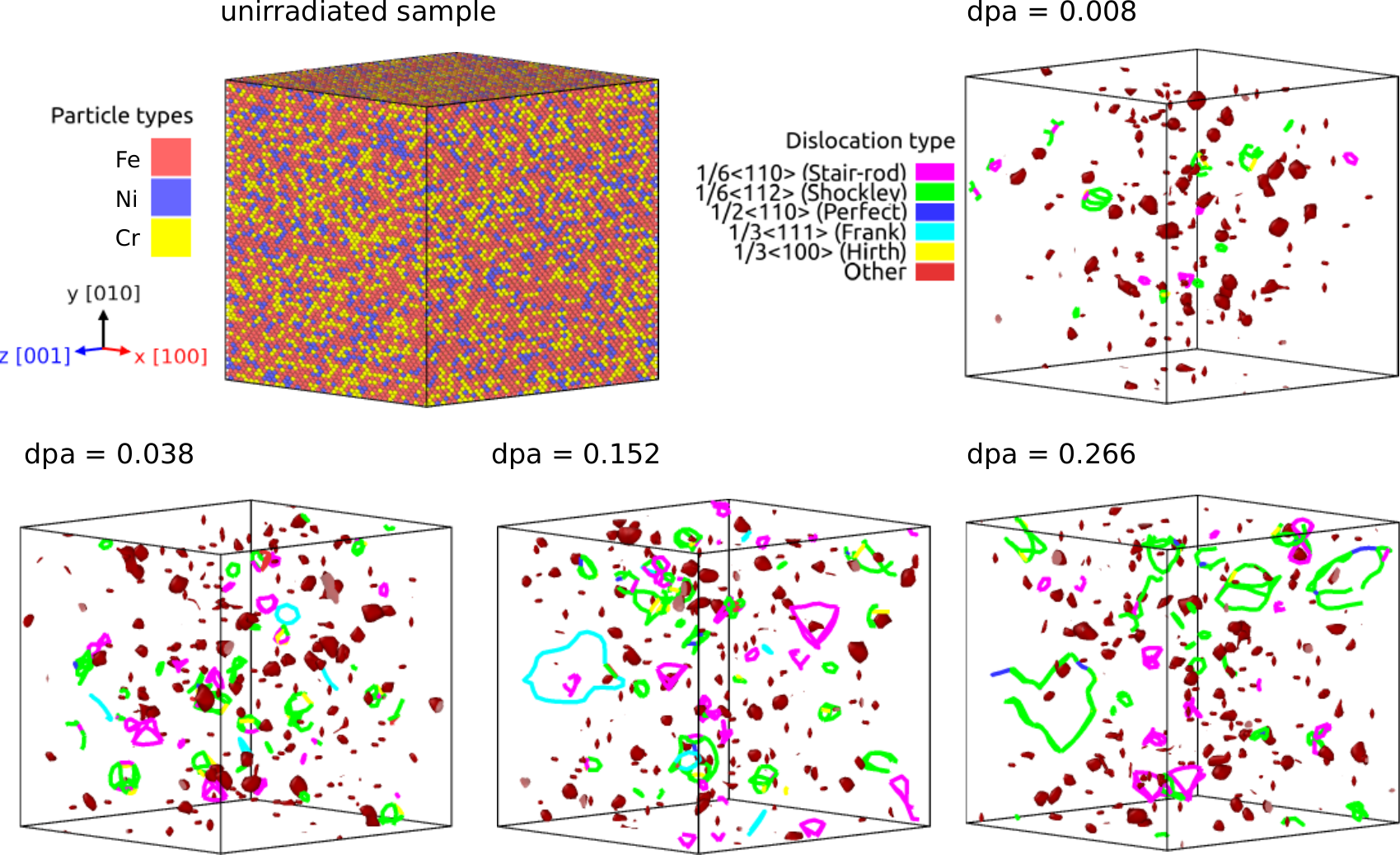}
    \caption{Structural evolution of radiation defects in Fe--Ni--Cr alloy with increasing dpa. The legend identifies the types of dislocation loops and their designations. Voids are represented in burgundy.}
    \label{fig:irradiated_samples}
\end{figure}

\textcolor{black}{To initiate the displacement cascade simulations, primary knock-on atoms (PKAs) were randomly selected from the simulation cell. Although the atomic masses of Fe, Cr, and Ni are relatively close, the initial velocity of each PKA was adjusted according to its mass to ensure a constant recoil energy of 10 keV, in accordance with the kinetic energy dependence on atomic mass. This approach allowed for a physically consistent treatment of recoil events across different atomic species.}
To achieve dpa values of 0.008, 0.038, 0.152, and 0.266, as shown in Fig. \ref{fig:irradiated_samples}; 20, 100, 400, and 700 consecutive recoil MD simulations were performed, respectively. 
Each recoil event was simulated for 20 ps under the NPT ensemble, followed by an additional 10 ps to allow for stress relaxation and swelling \citep{USTRZYCKA2024}. \textcolor{black}{Defect formation, including dislocation nucleation and void generation, was analyzed across various dpa levels using the Dislocation Extraction Algorithm (DXA) for dislocation analysis and Voronoi-based tessellation methods for identifying voids, as implemented in OVITO \citep{Stukowski_2010}.} \textcolor{black}{Dpa dose was calculated following the approach proposed by \citet[Eq. (10)]{Nordlund2018}, using the expression $\text{dpa} = \frac{1}{N} \cdot \frac{0.8 T_{\rm PKA}}{2E_d}$, where $T_{\rm PKA}$ denotes the kinetic energy of PKA and $N$  is the total number of atoms in the simulation cell. In this study, a constant value of displacement threshold energy $E_{\rm d}$ = 40\,\text{eV}  was used, which lies within the typical range reported for Fe, Ni, and Cr \citep{SELLAMI2019, ANDERSON2024}. While the model can accommodate element-specific $E_{\rm d}$ values, this approximation was adopted for consistency across the alloy system.}

Fig. \ref{fig:irradiated_samples} illustrates how microstructural features, including dislocation loops and void densities, evolve under increasing radiation exposure, highlighting the transition from isolated defects to a dense defect structure.
\textcolor{black}{Additional details on the development of radiation-induced defect structures in the samples are provided in Appendix~\ref{sec:appendixA}.}

\subsection{Simulations of fracture}
\label{sec: Section2.2}
\begin{figure}[b!]
\centering
\includegraphics[width=0.8\textwidth]{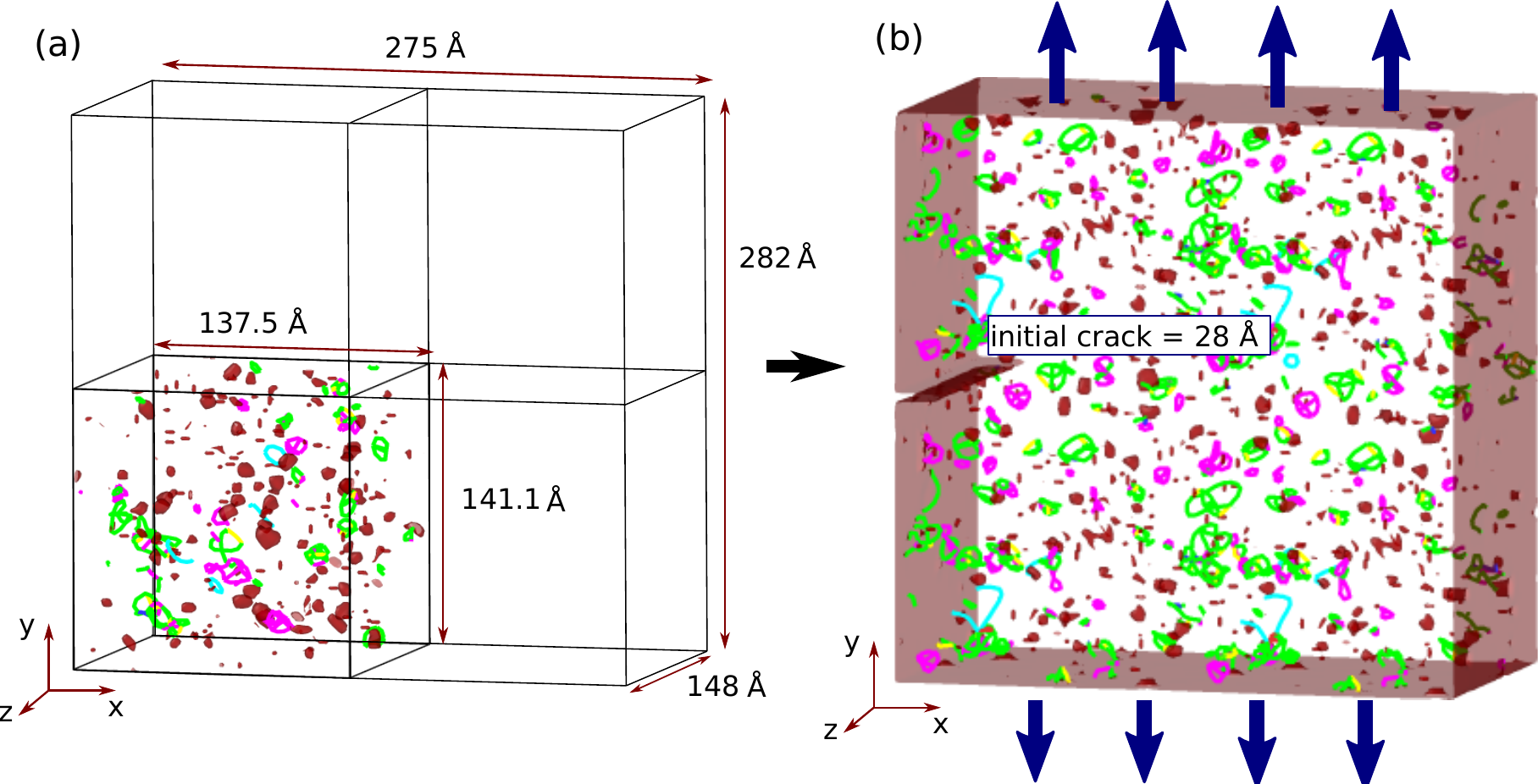}
\caption{Schematic representation of the 3D atomic-scale fracture model. (a) The simulation box prepared by replicating an irradiated sample along the $x$- and $y$-axes. (b) The model with an initial crack introduced at the midplane subjected to uniform elongation.}
    \label{fig:Crack_model}
\end{figure}

\vspace{0.3cm}
A three-dimensional (3D) atomic-scale model shown in Fig. \ref{fig:Crack_model} was prepared to enable the simulation and analysis of fracture mechanics phenomena. The irradiated samples obtained from the irradiation step (Section \ref{sec: Section2.1}) were duplicated using periodic replication techniques along the $x$- and $y$-axes to achieve the desired model dimensions of $275\times282\times148$ $\text{\AA}^3$. The simulation box, containing 1\;048\;320 atoms, provides a sufficient volume to capture crack-related phenomena without significant boundary interference. To ensure the stability of the fracture model under the specified conditions, a relaxation step was performed for 10 ps following an equilibration process using a canonical NPT ensemble for 100 ps at a target temperature of 300 K and zero stress. This equilibration phase allowed both atomic positions and simulation box dimensions to adjust dynamically, ensuring the system reached a stable thermodynamic state. Following equilibration, the NVT ensemble was applied to preserve a constant temperature throughout the deformation process, ensuring thermal stability and eliminating artifacts such as unintended energy dissipation or volume fluctuations that might influence crack propagation behavior. \textcolor{black}{An initial crack with a length of 28$\,\text{\AA}$ was introduced at the midplane of the specimen by selectively removing bonds between atoms along the $x$-axis over the full thickness of the specimen (i.e., across all atomic layers in the $z$-direction), creating a sharp discontinuity.}

\begin{figure}[t!]
\centering
\includegraphics[width=1\textwidth]{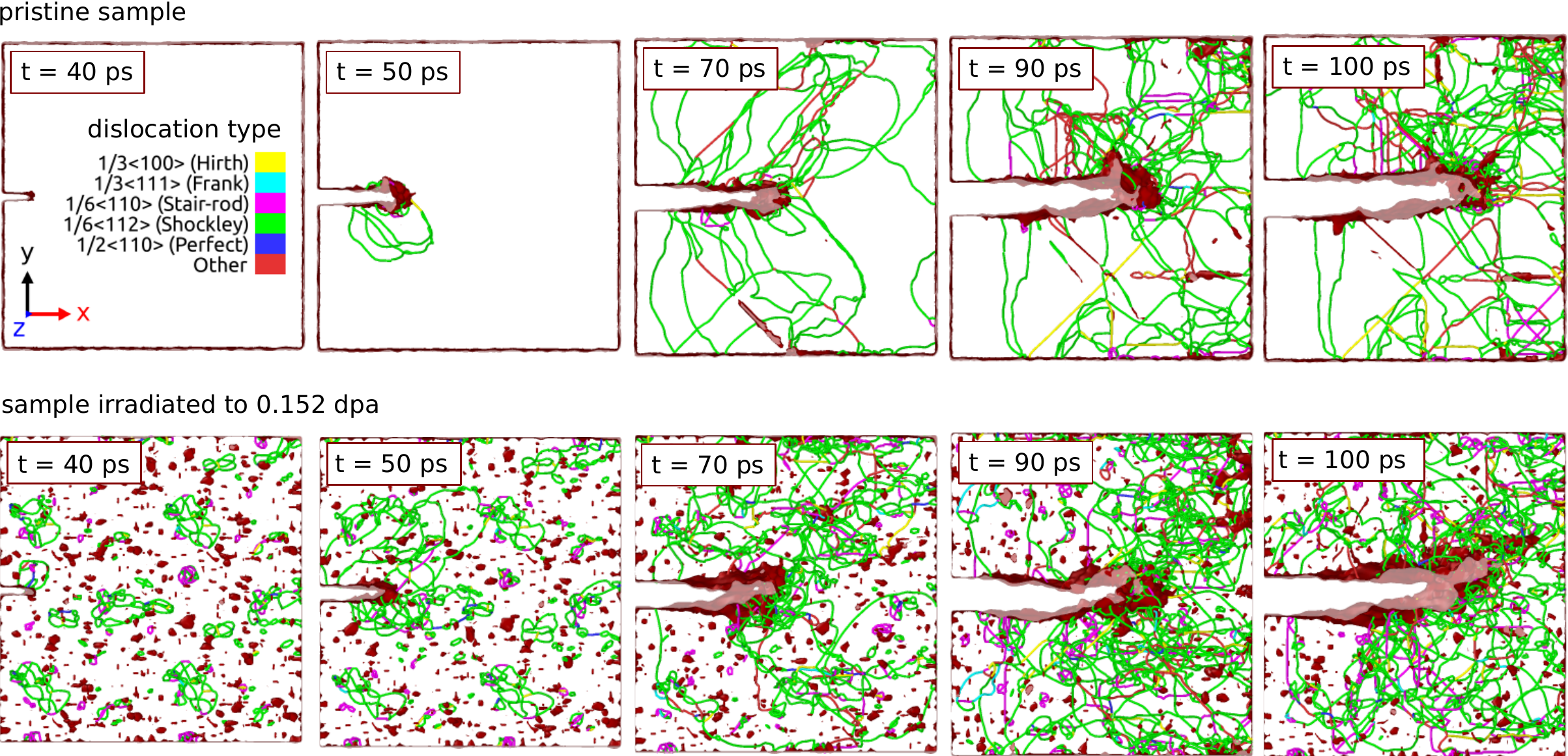}
   \caption{Crack propagation for non-irradiated (top) and irradiated (bottom) sample.}
    \label{fig: crack propagation_EAM11_300K}
\end{figure}
Due to the low thickness of the crystal compared to its height and width, this problem is modeled as a constrained 3D system \citep{Krull2011}. Periodic boundary conditions are applied along the thickness of the crystal (i.e., $z$-direction), while non-periodic boundary conditions are used in the $x$- and $y$-directions, corresponding to the crack propagation and loading direction, respectively. To prevent necking and minimize residual effects (unwanted structural deformations), the atoms at the left and right edges of the sample (see Fig. \ref{fig:Crack_model}) are restricted from moving in the $x$-direction but remain free to move along the $y$- and $z$-axes, in accordance with the assumed potential and loading conditions. In addition, two atomic layers at the top and bottom of the sample along the $y$-direction are kept frozen. The system is subjected to uniform elongation along the $y$-direction, applied to the upper and lower boundaries. The displacement rate is set to 0.08 $\text{\AA}$/{ps}, corresponding to a global strain rate of {$10^9$} 1/s. Following the criteria established by \citet{buehler2008}, a time step of 1 fs is used for the simulations. This approach provides the foundation for analyzing the mechanical response and fracture behavior of both unirradiated and irradiated samples. 

\textcolor{black}{A detailed analysis of alternative boundary condition configurations---fully periodic (ppp), fully free (ffp), and non-periodic shrink-wrapped (ssp)---is presented in Appendix~\ref{sec:appendixB}. This analysis evaluates their effect on plastic deformation and fracture behavior and provides the rationale for employing the constrained ssp setup in the main study.}

The crack propagation in both a unirradiated sample and in a selected irradiated sample (0.152 dpa) is illustrated in Fig.~\ref{fig: crack propagation_EAM11_300K}, where the role of radiation-induced defects on dislocation behavior and crack path characteristics is highlighted. In the pristine sample, crack growth generates significant dislocation activity near the crack tip, dissipating energy through plastic deformation and effectively blunting the crack tip. 
In the irradiated sample, radiation defects additionally increase the overall dislocation density in the whole volume. \textcolor{black}{Strain-field data further illustrating the differences between the deformation responses of unirradiated and irradiated materials is presented in Appendix~\ref{sec:appendixC}.} 

\section {Characterization of mechanisms governing crack propagation in irradiated materials}
\label{sec: Section3}
\vspace{0.3cm}

\textcolor{black}{In this section, various modes of crack propagation and defect interaction across different irradiation levels are examined in detail. Section~\ref{sec: Subection31} examines characteristic modes of crack growth, highlighting how radiation-induced defects alter crack trajectories and lead to phenomena such as crack branching, void coalescence, and secondary crack formation. Section~\ref{sec: Subection32} delves into the fundamental mechanisms governing defect--crack interactions, focusing on how \textcolor{black}{isolated defects} (e.g., voids, dislocation loops, rigid inclusions) either facilitate or impede crack propagation and how these interactions influence fracture behavior. Finally, defect-driven fracture mechanisms are summarized in Section~\ref{sec:33}.}

\subsection{Characteristic modes of crack growth in irradiated structures}
\label{sec: Subection31}
\vspace{0.3cm}
\begin{figure}[t!]
 \centering
\includegraphics[scale=0.4]{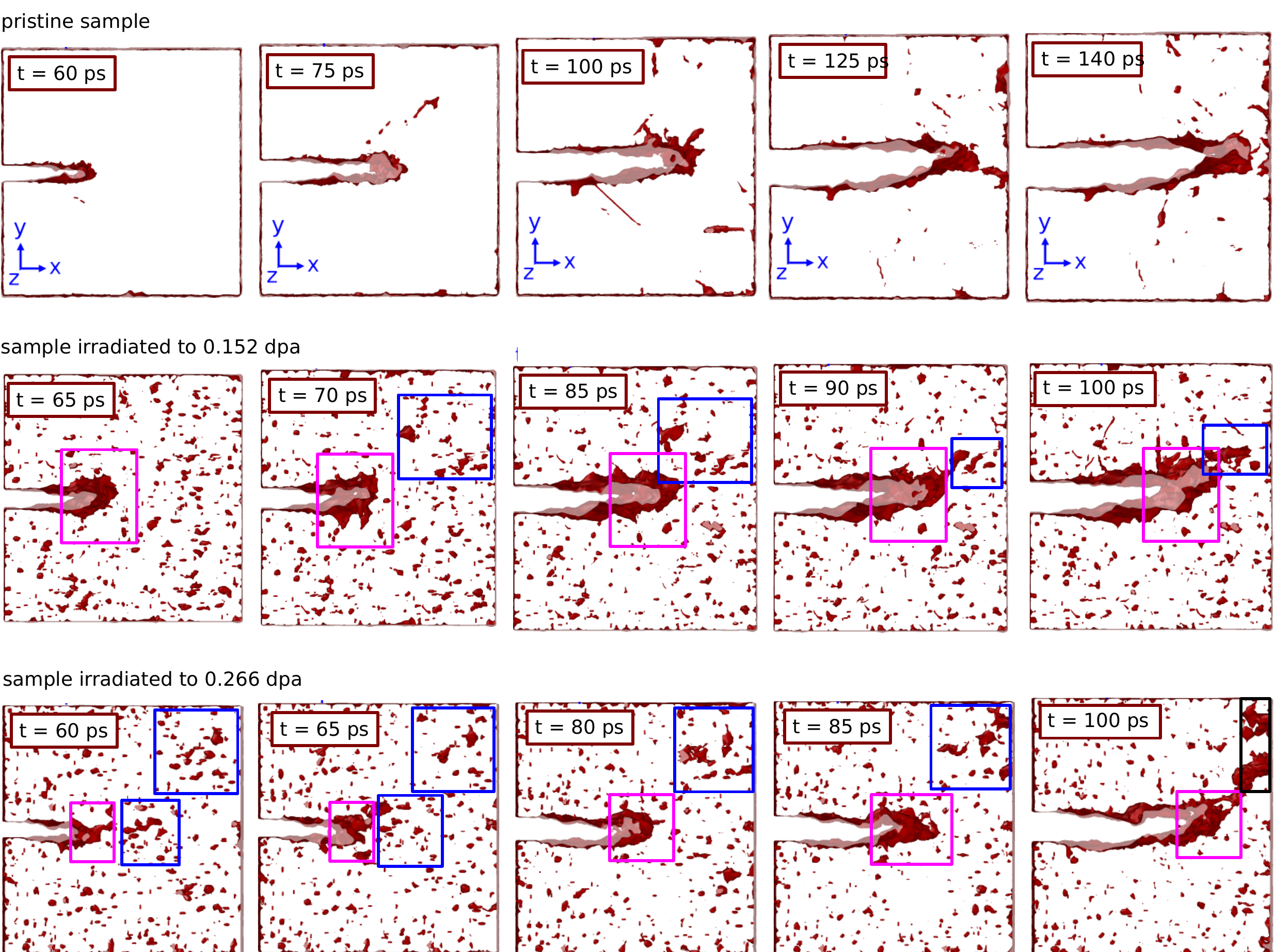} 
\caption{Influence of radiation-induced voids and dislocation obstacles on crack growth paths. The coalescence of voids is highlighted with a blue frame, the amorphous zone with a pink frame, and the formation of secondary cracks with a black frame.}
\label{fig: crack_coalescence}
\end{figure}
In this subsection, representative cases are examined to show the complex mechanisms driving crack evolution under irradiation. Fig. \ref{fig: crack_coalescence} illustrates the crack propagation process in both unirradiated and irradiated samples.

In the unirradiated sample (top row of Figs \ref{fig: crack propagation_EAM11_300K} and \ref{fig: crack_coalescence}), the crack advances with minimal resistance, exhibiting a relatively smooth growth path accompanied by a developed plastic zone at the crack tip. The emission of dislocations near the crack tip is observed to hinder crack propagation by promoting localized plastic deformation. This dislocation activity contributes to crack tip blunting (Fig. \ref{fig: crack_coalescence}: at 75 to 100\,ps), as evidenced by the gradual widening of the crack tip and the formation of a diffuse plastic zone (Fig. \ref{fig: crack propagation_EAM11_300K}: at 70 to 100\,ps). 

In the irradiated samples, the crack growth dynamics is distinctly different. 
\textcolor{black}{The sample with 0.152 dpa (Fig. \ref{fig: crack_coalescence}: middle row) shows a crack path that is restricted and/or redirected under the influence of radiation-induced defects. 
Unlike the unirradiated sample in Fig. \ref{fig: crack propagation_EAM11_300K} (top row, unirradiated sample), where dislocations are primarily emitted from the crack tip, the irradiated sample exhibits additional dislocation activity originating from radiation-induced defects (Fig. \ref{fig: crack propagation_EAM11_300K}, sample irradiated to 0.152 dpa).} The defects not only concentrate strain and contribute to higher dislocation density but also limit the mobility of the emitted dislocations, thereby restricting plastic deformation and modifying the crack’s trajectory.
As a result, an amorphous zone with a varying width of 2--5 nm is formed at the crack tip (in Fig. \ref{fig: crack_coalescence}, marked with a pink frame). The zone evolves dynamically, propagating together with the crack tip.
Furthermore, under mechanical loads, as the crack advances, radiation-induced voids expand and coalesce, forming larger void structures. The blue frames in Fig. \ref{fig: crack_coalescence} highlight instances where previously isolated voids join together. This coalescence is driven by localized plastic deformation, which facilitates void merging and provides an easier path for crack propagation. Notably, this mechanism is analogous to void coalescence observed in non-irradiated ductile materials \citep{TVERGAARD2011}. Near the crack tip, coalescing voids create additional weak zones, directing the crack toward areas of lower resistance. This process enhances crack branching and influences its overall morphology. Eventually, as the crack encounters clusters of coalesced voids, these merged structures integrate into the main crack, leading to more irregular and complex crack propagation patterns (Fig. \ref{fig: crack_coalescence}; at 100\,ps).

In the sample irradiated to 0.266 dpa (Fig. \ref{fig: crack_coalescence}; bottom row), the crack propagation is on the one hand blocked by the dense defect field, leading to the formation of an amorphous zone at the crack tip (marked with a pink frame), on the other hand, the crack propagation is promoted by the coalescence of voids at the crack tip (marked with a blue frame). In addition, a secondary crack begins to form on the opposite side (Fig.~\ref{fig: crack_coalescence};  at 100\,ps, marked with a black frame), initiating around radiation-induced voids. The development of this secondary crack is also driven by the coalescence of voids (marked with a blue frame), which create weak regions that facilitate crack initiation and propagation. Eventually, the primary and secondary cracks merge, resulting in a more complex crack path.

Fig. \ref{fig: crack_coalescence} reveals the dynamic interaction between crack progression and void evolution, demonstrating how void growth, void coalescence, and subsequent merging with the crack enhance crack propagation. Moreover, Fig. \ref{fig: crack_coalescence} discloses that in irradiated materials, secondary cracks can form if the stress level in the sample is sufficiently high and the primary crack is blocked by tangled dislocations or amorphous zones \citep{LI2024}. When radiation voids develop and coalesce in other regions of the material, these zones become weakened and may serve as initiation sites for secondary cracks. This process ultimately leads to a complex fracture morphology that is significantly influenced by void accumulation and absorption.

\begin{figure}[t!]
\centering
\includegraphics[scale=0.4]{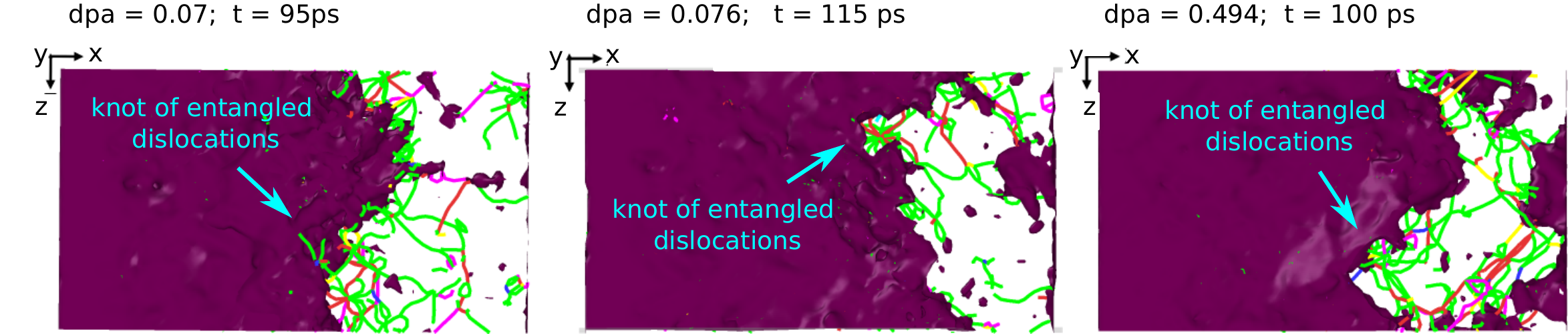}
\caption{Clustering of radiation-induced dislocation loops near the crack front. The burgundy color indicates a cross-section of the crack in the $x$-$z$ plane, highlighting a cross-sectional view of the crack.}
\label{fig: blocked crack}
\end{figure}

\textcolor{black}{Another critical mechanism that influences crack growth and alters fracture behavior is the interaction between the propagating crack and radiation-induced dislocation loops. This mechanism is illustrated in Fig.~\ref{fig: blocked crack} for different dpa levels. The dislocation loops create entangled structures that act as barriers to crack advance by reducing dislocation mobility near the crack tip. When the crack front encounters these obstacles, further propagation is hindered, and localized zones of increased resistance to fracture develop.}

\textcolor{black}{The above analysis highlights the contrasting roles of voids and dislocation loops in crack propagation. Voids facilitate crack advance by reducing local cohesion, while dislocation loops act as barriers that slow down or block the crack front. At higher dpa levels, these effects can coexist—crack branching may occur when the main crack is obstructed by loop accumulations, while secondary cracks initiate in weakened regions around radiation voids (see Figs. \ref{fig: crack_coalescence} and ~\ref{fig: blocked crack}). Together, these competing mechanisms contribute to the complex fracture morphology observed in irradiated materials.}
\subsection{Fundamental mechanisms governing defect--crack interactions}
\label{sec: Subection32}
\vspace{0.3cm}
\textcolor{black}{The previous subsection examined the overall effect of irradiation on crack dynamics and microstructure evolution. This section focuses on selected examples of how isolated, typical radiation-induced defects interact with the crack front.}

\begin{figure}[t!]
\centering
\includegraphics[scale=0.4]{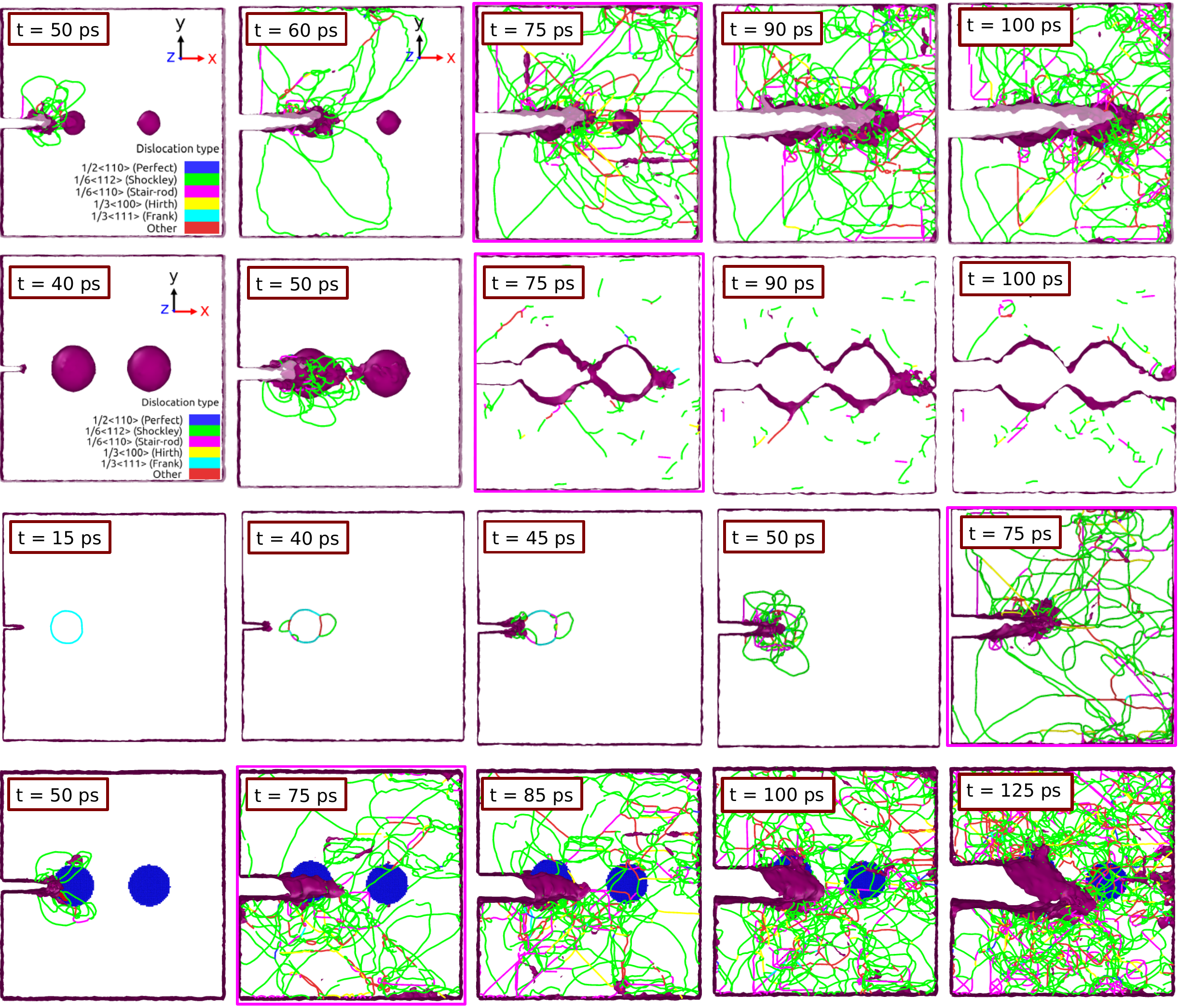}
\caption{Fundamental mechanisms of a single defect–crack interaction governing crack propagation. The individual rows correspond to small voids (top), large voids, dislocation loop, and rigid inclusion (bottom). Frames outlined in pink correspond to the same moment in time (75ps).}
\label{fig: physical mechanisms}
\end{figure}

\textcolor{black}{Fig.~\ref{fig: physical mechanisms} presents representative atomistic simulations showing the effect of specific defect types, small and large voids, dislocation loops, and rigid inclusions on crack propagation.} \textcolor{black}{Rigid inclusions are used here as a simplified model to mimic the strong barrier effect of highly entangled dislocation structures.} \textcolor{black}{Each case illustrates a distinct interaction mechanism and its influence on the crack path and morphology. 
This approach allows to isolate and better understand the fundamental defect–crack interaction mechanisms that contribute to the macroscopic fracture behavior of irradiated alloys.}

To investigate these interactions, simulations were conducted using a sample with dimensions of  $275\times282\times148$ $\text{\AA}^3$, consistent with the setup described in Section \ref{sec: Section2.2}. The defects were designed to reflect their typical geometries as observed in experiments: small voids with a diameter of 2.5 nm, large voids with a diameter of 5 nm, dislocation loops spanning 5 nm, and rigid inclusions with diameter of 5 nm. These controlled configurations enabled a systematic exploration of the effects of defect size and type on crack propagation.
In the first row of Fig. \ref{fig: physical mechanisms}, the crack approaches two small voids, each with a diameter of 2.5 nm. These voids act as localized weak points in the material, reducing the effective cohesive energy and creating regions of lower fracture resistance. As the crack propagates, it preferentially connects to these voids, accelerating crack growth. This mechanism highlights how voids act as stress concentrators, lowering the energy barrier for crack propagation and promoting fracture along paths of least resistance \citep{bitzek2008, chandra2016, li2017}.

In the second row of Fig. \ref{fig: physical mechanisms}, the crack approaches two larger voids, each with a diameter of 5 nm. Due to the increased size, these voids have a more pronounced effect on crack propagation, creating a disruption in the crack's path as it merges with the voids. This interaction accelerates the crack propagation as the crack incorporates the void regions into its path.

In the third row of Fig. \ref{fig: physical mechanisms}, the crack interacts with a dislocation loop spanning 5 nm. This interaction is distinct because the dislocation loop acts as a barrier, impeding crack growth by obstructing dislocation motion. As the crack encounters the dislocation loop, it experiences increased resistance, which can alter its trajectory or even temporarily halt its progression. 

In the fourth row of Fig. \ref{fig: physical mechanisms}, the crack faces two rigid inclusions, each with dimensions of 5 nm. The rigid inclusions are introduced through the application of boundary conditions that constrain any deformation within the inclusion region during the simulation, thereby ensuring the inclusion remains a fixed, immovable obstacle within the material. These inclusions represent strong obstacles that divert the crack path, leading to a more tortuous and irregular morphology. Unlike voids, which weaken the material locally, rigid inclusions increase local resistance and force the crack to change direction. This can result in additional dislocation activity around the inclusions as the crack adapts to the path of least resistance. This results in a more tortuous and branched crack morphology.

\textcolor{black}{Fig.~\ref{fig: physical mechanisms} shows how the crack response depends on defect type and size. This analysis confirms that voids act as local weakening sites that promote crack growth, while dislocation loops and rigid inclusions impede propagation by increasing resistance near the crack tip. The transition from facilitation to obstruction highlights the complex role of microstructural features in shaping crack morphology.}

\textcolor{black}{These insights complement the broader findings from irradiated samples discussed in the previous subsection. While Section~\ref{sec: Subection31} explored collective crack behavior across varying irradiation levels, the current analysis in Section~\ref{sec: Subection32} isolates individual mechanisms to reveal their distinct contributions to fracture evolution in irradiated alloys.}
\subsection{Summary of defect-driven fracture mechanisms in irradiated materials}
\label{sec:33}
\vspace{0.3cm}
This section presents a detailed summary of the mechanisms governing crack propagation in irradiated materials, as revealed by the MD simulations reported above, emphasizing the complex interactions between radiation-induced defects and fracture dynamics. The results highlight the complexity of mechanisms that dictate fracture behavior,  providing fundamental insights into the competing factors of crack growth. The key findings are summarized as follows:
\begin{itemize}
\item \emph{Mechanism of void--crack interactions:~} Voids act as local weak points in the material, reducing the effective energy and creating pathways of reduced resistance for crack growth. In regions with high void concentrations, the crack propagates more easily, as the voids lower the local stress required for atomic bond breaking. This mechanism is particularly pronounced when voids coalesce ahead of the crack tip, forming larger voids that further weaken the material. The coalescence of voids not only accelerates crack propagation but also leads to irregular crack morphologies, as the crack tends to follow the path of least resistance through these weakened zones. This process often results in crack branching, where the primary crack splits into multiple secondary cracks, increasing the overall fracture surface area and contributing to a more complex fracture pattern. \textcolor{black}{Crack--void interaction mechanisms with artificially introduced defects were previously studied using molecular dynamics in single-crystal silicon \citep{WANG2015}, where the effect of void spacing was analyzed. Similar simulations in aluminum \citep{chandra2016} showed that vacancy clusters can alter crack tip behavior by modifying stress concentrations and promoting crack tip blunting.}
\item \emph{Mechanism of dislocation loop--crack interactions:~} In contrast to voids, dislocation loops act as barriers to crack propagation. These loops, formed from clusters of interstitial atoms that form stable structures, obstruct the motion of dislocations near the crack tip, creating localized stress concentrations. When the crack encounters a dislocation loop, its progression is temporarily halted. This interaction can lead to the emission of additional dislocations from the crack tip, promoting localized plastic deformation and energy dissipation. However, in irradiated materials, the mobility of these dislocations is significantly restricted by the presence of other radiation-induced defects, such as voids and additional dislocation loops. As a result, the energy dissipation through plasticity is reduced, and the material becomes more prone to brittle fracture. Notably, dislocation loops, due to their immobile nature, can also act similarly to rigid inclusions by deflecting crack paths, increasing local stress concentrations, and altering crack propagation trajectories. \textcolor{black}{Dislocation--crack interactions have also been investigated in earlier atomistic studies on ideal crystalline systems \citep{SHENG2023a,SHENG2023b}. MD simulations in nickel showed that dislocations approaching the crack tip can trigger dislocation emission, cross-slip, and the development of a localized plastic zone \citep{bitzek2008}. However, these studies did not consider radiation-induced dislocation loops or their evolution, and therefore do not capture the complex mechanisms present in irradiated materials.}
\item \emph{Radiation-induced defects as nucleation sites for dislocations:~} In addition to acting as obstacles, irradiation-induced defects serve as preferential nucleation sites for dislocation formation. Defect clusters, such as voids and dislocation loops, introduce local stress concentrations that facilitate dislocation emission. As irradiation levels increase, dislocation nucleation leads to reduced plasticity and a more brittle fracture response. This dual role of defects, both as dislocation sources and barriers, critically influences the balance between ductile and brittle failure mechanisms. 
\textcolor{black}{This behavior was also observed in our study through atomistic simulations of shear deformation in irradiated Fe--Ni--Cr alloys, where dislocation nucleation was observed at radiation-induced defect clusters \citep{USTRZYCKA2024}.}
\item \emph{Mechanism of amorphous zone development at the crack tip:~} The amorphous zones which consist of disordered atomic structures form as a result of the accumulation of radiation-induced defects, such as dislocation loops. These zones, act as barriers to crack propagation, absorbing energy and temporarily halting the crack's advance. However, these zones also contribute to localized embrittlement, as the disordered structure reduces the material's ability to deform plastically. As a result, the secondary cracks can be formed, further increasing the complexity of the fracture process.
\textcolor{black}{Similar amorphization-induced crack shielding has also been reported in Si single crystals under high contact stresses during grinding, where amorphous zones were shown to suppress cracking by altering the local stress state~\citep{LI2024}.}
\item \emph{Crack branching and secondary crack formation:~} High defect densities, particularly clusters of dislocation loops, lead to crack branching and the initiation of secondary cracks. When the primary crack encounters dense defect fields, it can branch or form secondary cracks, which eventually coalesce with the main crack. This results in irregular fracture patterns and an increased fracture surface area. The formation of secondary cracks is driven by the stress concentrations around radiation-induced defects, which provide initiation sites for new cracks.
\textcolor{black}{Crack branching and secondary cracks are often observed in MD simulations for brittle materials, e.g.~\citep{nakano1995,luo2021}, but also for ductile materials, e.g., for titanium under cyclic loading~\citep{liu2024}.}
\end{itemize}

The fracture behavior of irradiated materials is governed by a complex and interconnected interplay of mechanisms, including void--crack interactions, dislocation loop--crack interactions, rigid inclusions, amorphous zone formation, void coalescence, crack branching, and stress concentration effects. These mechanisms are not isolated; they are often coupled and interlocked in specific regions and under certain conditions, creating a highly dynamic and interdependent fracture process.

By examining each mechanism individually, a clearer understanding of its specific role in the fracture process can be obtained. However, it is only through the collective analysis of these mechanisms that the complexity of fracture dynamics in irradiated materials can be fully comprehended. Together, these mechanisms determine the crack trajectory, propagation velocity, and overall fracture morphology. The material's ability to dissipate energy through plasticity is significantly altered by the presence of radiation-induced defects, leading to a transition from ductile to brittle fracture. This transition is driven by the combined effects of defect interactions, which reduce the material's capacity for plastic deformation and promote brittle failure modes.

\section{Analysis of the critical strain energy release rate}
\label{sec: Section4}
\textcolor{black}{This section establishes a theoretical and computational framework for determiening the critical strain energy release rate ($G_{\rm c}$) in irradiated materials, bridging classical fracture mechanics with atomistic simulations. It examines the interplay between bond rupture (surface energy, $\gamma_{\rm s}$) and dislocation-mediated plasticity (plastic work, $w_{\rm p}$) under irradiation, using molecular dynamics to isolate irradiation- and temperature-dependent contributions to $G_{\rm c}$. }
\subsection{Framework for energy-based fracture analysis}
\label{sec: Section4.1}
\vspace{0.3cm}
In brittle fracture, the energy expended in the fracture process primarily goes into creating two new surfaces, where atomic bonds are broken. The energy needed for crack growth to overcome the cohesive forces between atoms corresponds to increased surface energy supplied by the elastic strain energy stored in the material. The classical Griffith theory defines the critical energy release rate as $G_{\rm c}\!\!=\!2\gamma_{\rm s}$, where $\gamma_{\rm s}$ represents the fracture surface energy, assuming purely brittle propagation with energy dissipation confined to the creation of new free surfaces.

However, in materials where crack extension is accompanied by plastic deformation, the Griffith framework becomes insufficient. The works by \citet{orowan1945notch}, \citet{JOKL1980} and \citet{BELTZ1992} established that energy dissipation during fracture involves both bond-breaking and plastic work $w_{\rm p}$ from dislocation nucleation and slip. \citet{JOKL1980} derived a thermodynamic criterion reformulating the critical energy release rate as:
\begin{equation}
G_{\rm c} = 2\gamma_{\rm s} + w_{\rm p},
\label{eq:Gc }
\end{equation}
where $\gamma_{\rm s}$ represents the fracture surface energy and $w_{\rm p}$ is the plastic work per unit crack area due to dislocation activity. Concurrently, \citet{BELTZ1992} analyzed the competition between cleavage and dislocation nucleation at crack tips, introducing the concept of an unstable stacking energy as the barrier for dislocation emission. Together, these works provided a foundation for computing $G_{\rm c}$ in materials where plasticity and defects govern fracture.

Recent advances by \citet{XU2020}, see also \citet{BARIK2024}, have applied this framework to atomistic simulations, directly calculating the critical energy release rate $G_{\rm c}$ for systems with crack-tip plasticity. It has been shown that $w_{\rm p}$ depends on the sample size, which is strongly limited in MD due to computational cost. In this approach, $G_{\rm c}$ is defined as the increment of the total dissipated energy, $\mathrm{d}E_{\rm dis}$, per increment of the projected crack surface area,  $\mathrm{d}A_{\rm crack}$, thus
\begin{equation}
G_{\rm c} = \frac{\mathrm{\partial}E_{\rm dis}}{\mathrm{\partial}A_{\rm crack}} \approx \frac{\mathrm{d}W\!-\mathrm{d}U}{\mathrm{d} A_{\rm crack}}.
\label{eq:Edis}
\end{equation}
Here, referring to the MD simulations, the dissipated energy increment $\mathrm{d} E_{\rm dis} \approx \mathrm{d} W - \mathrm{d} U$ is determined as the difference between the increments of the external work $W$ (derived from the stress--strain response) and internal energy $U$ (comprising potential and kinetic energies). Note that, under isothermal conditions (maintained by the thermostat), the dissipated energy corresponds to the change in Helmholtz free energy, $\mathrm{d} E_{\rm dis} = \mathrm{d} W \!- \mathrm{d} F$,  where  $\mathrm{d} F = \mathrm{d} U - T \mathrm{d} S$, which includes contributions from both internal energy $U$ and entropy $S$. Since the entropy and its changes cannot be computed for the problem at hand, the entropy contribution is disregarded above so that $\mathrm{d} U$ is assumed to approximate $\mathrm{d} F$, thus $\mathrm{d} F \approx \mathrm{d} U$, as tacitly assumed in~\citep{XU2020, BARIK2024}. As a result, Eq.~\eqref{eq:Edis} is obtained, and thus the heat generated during the inelastic deformation process (and taken from the system by the thermostat to maintain a constant temperature) is adopted as a measure of the dissipated energy. Note that it is commonly accepted that the majority of plastic work is converted to heat and thus dissipated (the rest being stored in the material), although in the context of MD simulations this depends on the material, the stage of the deformation process, and whether the integral or instantaneous (differential) energy conversion rate is considered~\citep{Kositski2021, Stimac2022}. Nevertheless, the generated heat constitutes the main contribution in the energy balance, which justifies the above approximation.


\subsection{Effects of radiation on fracture energy}
\label{sec: Section4.2}
\vspace{0.3cm}
Irradiation-induced defects fundamentally alter fracture behavior by degrading atomic cohesion (which influences surface energy, $\gamma_{\rm s}$) and suppressing dislocation-mediated plasticity (which impacts plastic work, $w_{\rm p}$). The spatial distribution, size, and density of these defects govern their collective impact on fracture resistance, creating a microstructure-dependent competition between brittle crack propagation and energy absorption through plastic deformation.

The critical energy release rate $ G_{\rm c}$ is therefore explicitly controlled by the radiation dose (dpa), which specifies defect density, with temperature $T$ serving here as a secondary parameter to isolate thermal activation of plasticity, thus:
\begin{equation}
G_{\rm c}\! \left(\mathrm{dpa}, T\right) = 2\gamma_{\rm s}\left(\mathrm{dpa}, T\right) + w_{\rm p}\left(\mathrm{dpa}, T\right).
\label{eq:condensed}
\end{equation}
Note that temperature modulates $w_{\rm p}$ through thermally activated dislocation glide, a defining feature of temperature-driven DBT. In contrast, the irradiation-driven DBT investigated here is dominated by radiation-induced defects that inhibit dislocation mobility and thereby fundamentally alter $w_{\rm p}$.

To estimate the intrinsic surface energy $\gamma_{\rm s}(\mathrm{dpa}, T)$, first the separation energy $2\gamma_{\rm sep}(\mathrm{dpa}, T)$ is considered, as a lower bound:
\begin{equation}
2\gamma_{\rm s}\left(\mathrm{dpa}, T\right) \geq 2\gamma_{\rm {sep}}\left(\mathrm{dpa},T \right).
\label{eq: 2gamma}
\end{equation}
Here, $\gamma_{\rm sep}$ represents a lower estimate of the surface energy because it corresponds to an idealized surface created under perfect conditions rather than the actual surface resulting from the fracture process.

An upper bound for $\gamma_{\rm s}$ can be obtained independently from the critical energy release rate at absolute zero temperature ($T$ = 0K), where thermal activation is suppressed and plastic dissipation is limited ($w_{\rm p} \approx 0$):
\begin{equation}
2\gamma_{\rm {s}}\left(\mathrm{dpa}, T\right) \leq G_{\rm c}\! \left(\mathrm{dpa}, 0\rm K\right).
\label{eq: 2gamma_1}
\end{equation}

Finally, at finite temperatures ($T$>0K), the temperature-dependent plastic work term, $w_{\rm p}(\mathrm{dpa}, T)$, can be estimated as: 
\begin{equation} 
w_{\rm p}\left(\mathrm{dpa}, T\right) = G_{\rm c} \left(\mathrm{dpa}, T\right) - 2\gamma_{\rm s}\left(\mathrm{dpa}, T\right), \quad 2\gamma_{\rm {sep}}\left(\mathrm{dpa}, T\right) \leq 2\gamma_{\rm {s}}\left(\mathrm{dpa}, T\right) \leq G_{\rm c} \left(\mathrm{dpa}, 0{\rm K}\right).
\label{eq:wp_est_up} 
\end{equation}
Eq.~\eqref{eq:wp_est_up} isolates the energy dissipated by dislocations as they interact with radiation-induced defects, providing a direct measure of the material’s plastic response to irradiation.

The radiation-induced 
DBT can be characterized by a decrease in $w_{\rm p}(\mathrm{dpa}, T)$ as the radiation dose (dpa) increases. Radiation-induced defects affect dislocation motion, altering the balance between plasticity and bond rupture during crack propagation (see Section \ref{sec:33}). The following regimes can be identified based on the value of $w_{\rm p}(\mathrm{dpa}, T)$:
\begin{itemize} 
\item \textit{high $w_{\rm p}(\mathrm{dpa}, T)$ -- ductile regime:~} plasticity dominates, and dislocations play a crucial role in shielding the crack tip; 
\item \textit{low $w_{\rm p}(\mathrm{dpa}, T)$ -- brittle regime:~} defect-induced hardening significantly restricts dislocation motion, leaving the crack tip unshielded and causing crack propagation to occur primarily through bond rupture, leading to a more brittle fracture response. 
\end{itemize}
\subsection{Energy dissipation in irradiated materials}
\label{sec: Section4.3}
\vspace{0.3cm}
In this subsection, the mechanisms of energy dissipation during crack propagation in irradiated materials are investigated across varying radiation doses, drawing on the MD simulation results presented in Section~\ref{sec: Section3}. By analyzing the interplay between internal energy $U$, external work $W$, and stress--strain behavior, the influence of defect populations on the critical energy release rate $G_{\rm c}$ and the balance between surface energy $\gamma_{\rm s}$ and plastic work dissipation $w_{\rm p}$ is quantified.

In order to evaluate the plastic work $w_{\rm p}(\mathrm{dpa}, T)$ as defined in Eq. \eqref{eq:wp_est_up}, the critical energy release rate $G_{\rm c}\!\left(\mathrm{dpa}, 0\rm K\right)$ should be determined at absolute zero temperature. 
To pragmatically approximate $G_{\rm c}\!\left(\mathrm{dpa}, 0\rm K\right)$, a finite simulation temperature of $T$=10K is employed,
\begin{equation} 
G_{\rm c}\! \left(\mathrm{dpa}, 10 \rm K\right) \approx G_{\rm c}\! \left(\mathrm{dpa}, 0\rm K\right). 
\label{eq: 10K} 
\end{equation}
This temperature is sufficiently low to suppress thermally activated processes,
thereby preserving the near-brittle fracture conditions required for isolating the surface energy contribution according to Eq. \eqref{eq:wp_est_up}. Thus, by evaluating fracture energy at two distinct temperature levels, the contribution of plastic work to energy dissipation can be effectively isolated. The low-temperature condition serves as a near-brittle reference state, minimizing dislocation activity and allowing the surface energy component to be approximated. In contrast, at higher temperatures, where plastic deformation mechanisms are active, the additional energy dissipation due to dislocation motion can be quantified.

\begin{figure}[t!]
\centering
\includegraphics[scale=0.4]{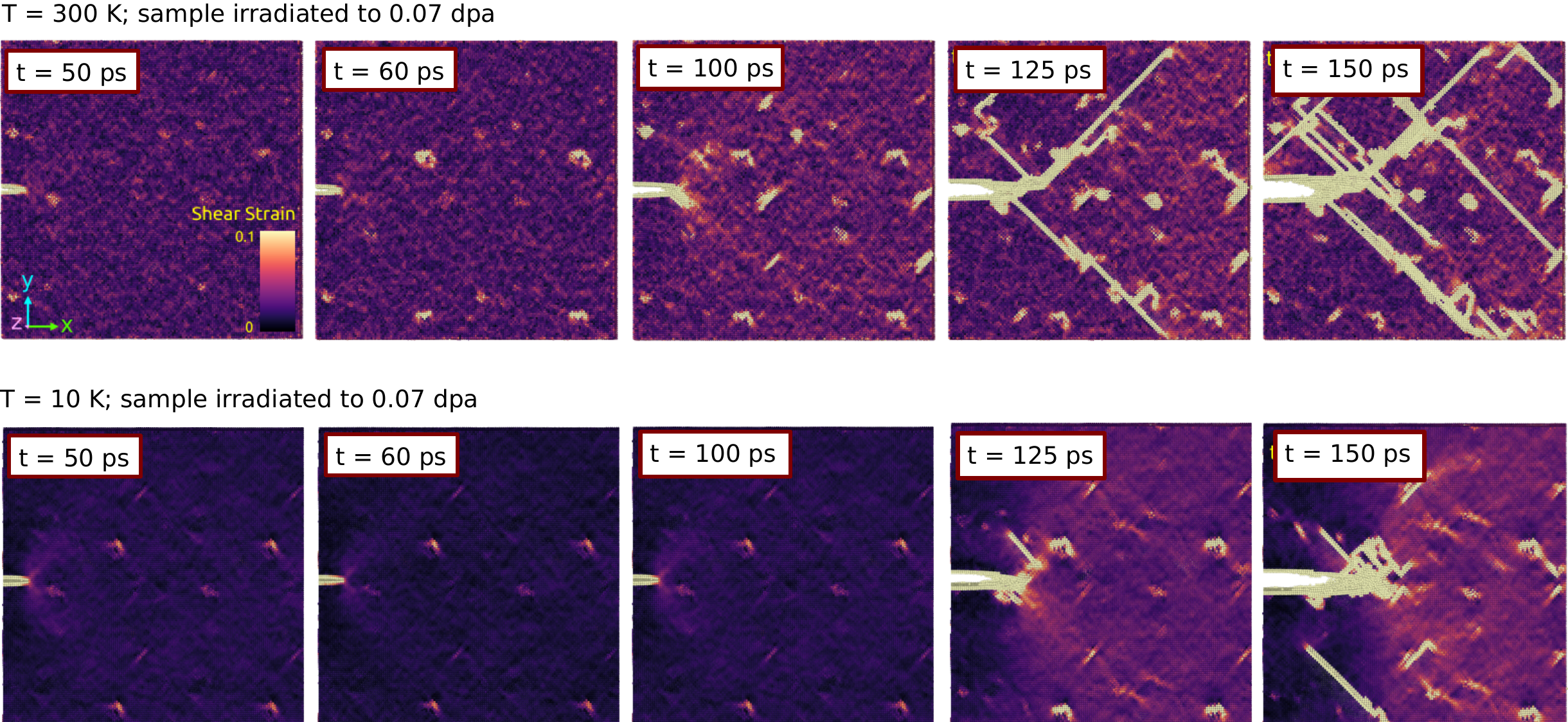}
\caption{\textcolor{black}{Evolution of the equivalent shear strain (von Mises invariant of the Green--Lagrangian strain tensor~\citep{Stukowski_2010})} in a selected cross-section of the samples irradiated to 0.07 dpa at 300K (top) and 10K (bottom).}
\label{fig: strain_300-10}
\end{figure}

Fig. \ref{fig: strain_300-10} compares the equivalent shear strain fields in cross-sections of samples irradiated to 0.07 dpa at 300K and 10K. At 300K (top row), the strain field exhibits a more widespread distribution than at 10K (bottom row), with deformation occurring throughout the sample and strain localizing particularly around the crack and radiation-induced defects. This reflects thermally activated dislocation motion, which enables plasticity and contributes to energy dissipation. In contrast, at 10K (bottom row), the strain is highly concentrated around the crack tip and radiation defects, with minimal deformation in the surrounding material. The suppression of strain outside these localized regions indicates restricted dislocation mobility and reduced plasticity, leading to more brittle fracture behavior. This comparison demonstrates that separation energy $\gamma_{\rm sep}$ and simulations at 10K can serve as a near-brittle reference state for approximating  $G_{\rm c}$(dpa, 0K) in Eq. \eqref{eq:wp_est_up}.

The effectiveness of this framework is demonstrated below through a detailed examination of energy dissipation characteristics in irradiated materials. 

\begin{figure}[t!]
\centering
\includegraphics[scale=0.31]{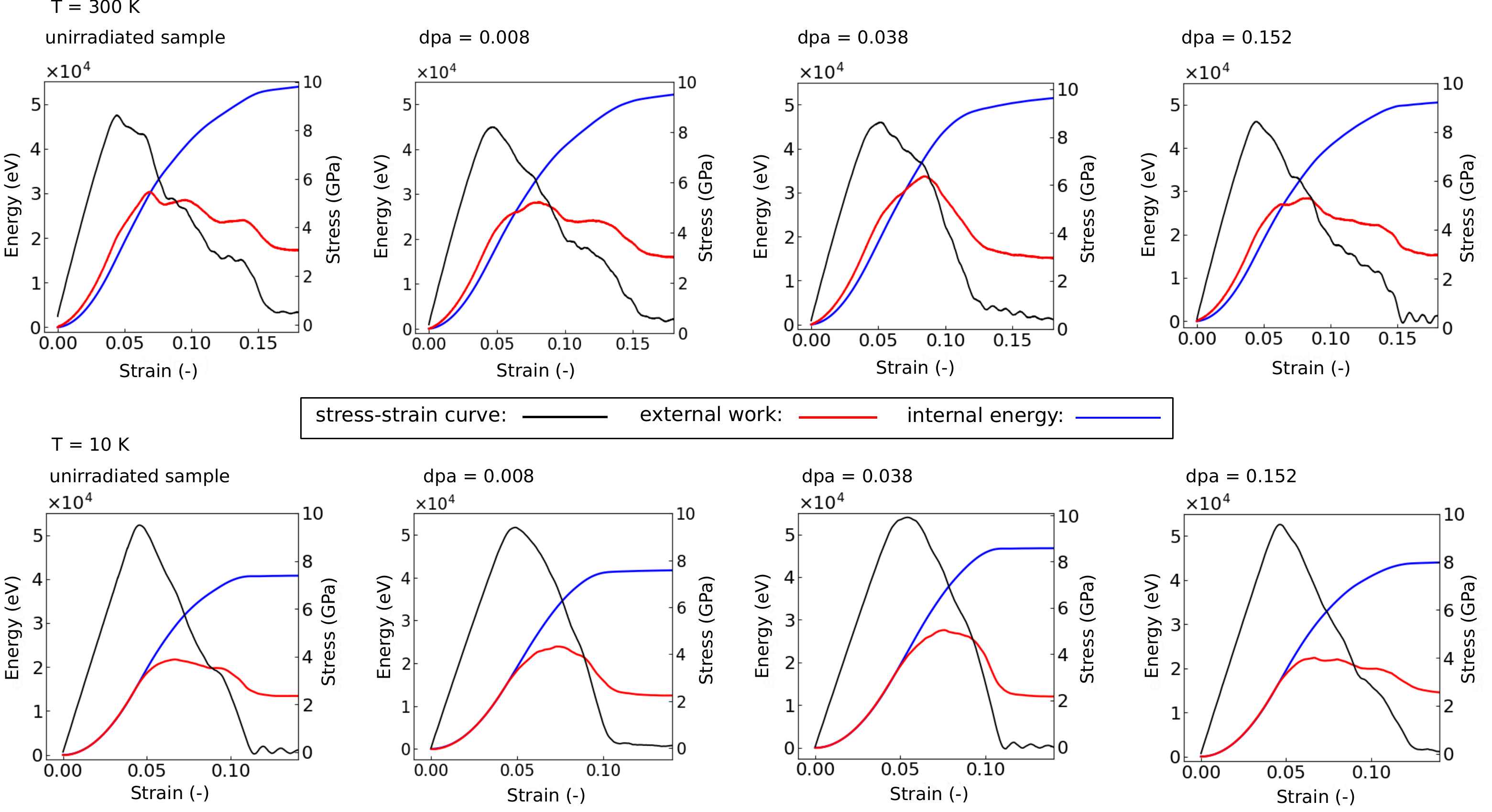}
\caption{Stress--strain curves and energy evolution for different irradiation levels at 300 K (top) and 10 K (bottom), showing internal energy $\Delta U$ (red), external work $W$ (blue), and stress--strain response (black).}
\label{fig: ss and energy}
\end{figure}

The evolution of internal energy $\Delta U=U-U_0$ (red curve) with $U_0$ referring to the initial reference state, external work $W$ (blue curve), and stress--strain response (black curve) for unirradiated and irradiated samples (0.008, 0.038, and 0.152 dpa) at two different temperatures, 300K and 10K are shown in Fig. \ref{fig: ss and energy}.
\begin{figure}[b!]
\centering
\includegraphics[scale=0.5]{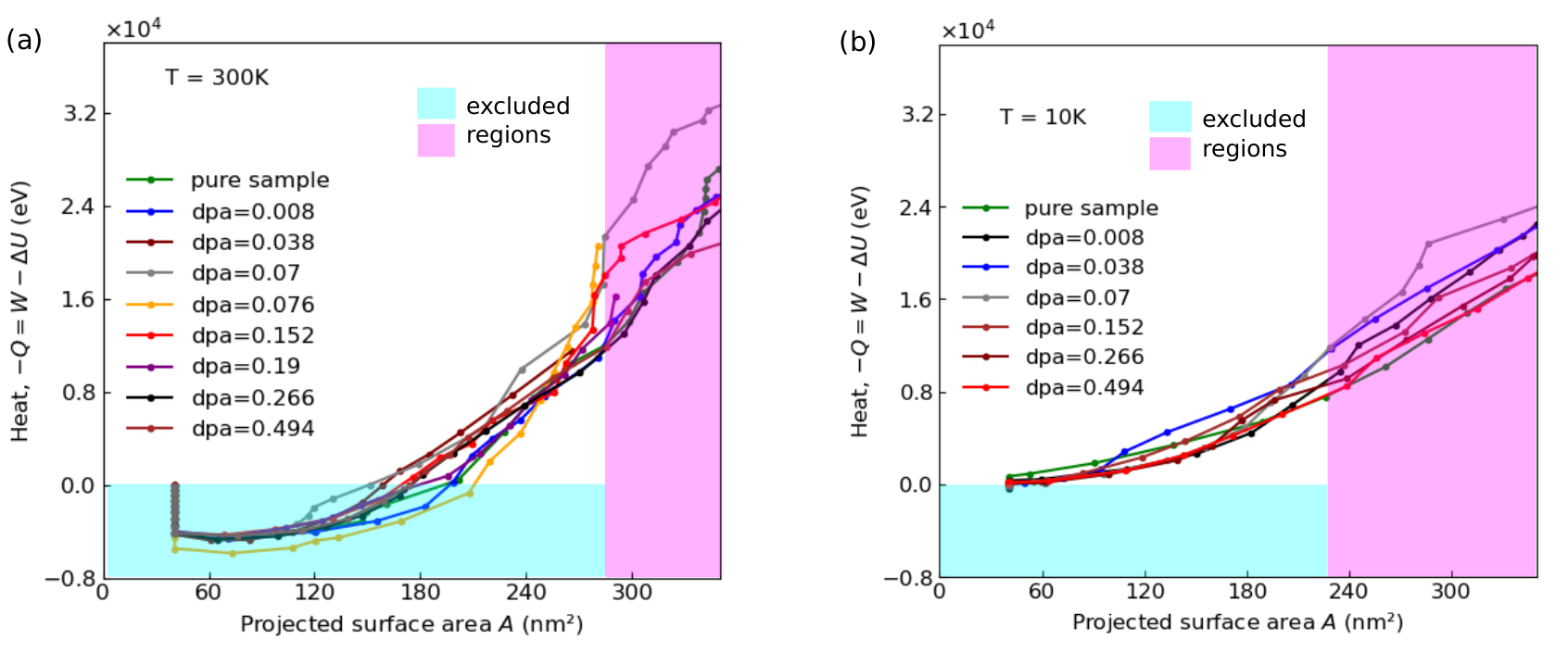}
\caption{\textcolor{black}{Heat taken from the system $(-Q = W - \Delta U)$ versus projected crack surface area $A$ for various dpa levels at (a) 300 K and (b) 10 K. Blue and pink shaded regions denote excluded intervals corresponding to initial plastic zone development and final boundary effects, respectively. The initial drop in $-Q$ observed at 300 K (but absent at 10 K) corresponds to the elastic deformation stage.}}
\label{fig: Ediss}
\end{figure}
At 300K, all samples exhibit discrete steps in their post-peak stress--strain curves, corresponding to crack arrest at the knots of entangled dislocations (compare Fig.~\ref{fig: blocked crack} and Section~\ref{sec: Section3}). 
These steps are also evident in the internal energy curve, reflecting energy relocation (or redistribution) as it diverts (or shifts) from bond rupture to dislocation nucleation.
The unirradiated sample displays a characteristic post-yield plateau (Fig.~\ref{fig: ss and energy}, top-left), indicative of homogeneous plastic zone development through unimpeded dislocation motion. Irradiation disrupts this plasticity: especially visible at 0.008 dpa and 0.152 dpa, the stress drops abruptly after yielding, indicating defect-dominated fracture driven by crack-void interactions with limited plastic dissipation (compare Section~\ref{sec: Section3}).

It is worth noting that the discrepancy between internal energy $\Delta U$ and external work $W$ in the elastic deformation regime, where $\Delta U$ marginally exceeds $W$, arises from the thermostat’s enforcement of isothermal conditions under the NVT ensemble. In fact, from the energy balance, the difference, $Q=\Delta U-W$, is the heat supplied to the system by the thermostat, which is positive ($Q>0$) in the elastic deformation regime. Once inelastic deformation mechanisms initiate (for the overall strain exceeding $\sim0.04$), the heat $Q$ becomes negative ($Q<0$), i.e., heat is taken from the system by the thermostat.

At 10K, the energy profiles and stress--strain behavior display a distinctly different response compared to $300\rm \,K$, reflecting the reduced atomic mobility and increased brittleness at low temperatures. In particular, at this temperature, internal energy closely follows external work during the elastic deformation stage and the heat supplied to the system is then negligible.

Fig. \ref{fig: Ediss} shows the heat taken from the system ($-Q=W-\Delta U$)
as a function of the projected crack surface area $A$ for all samples considered. Here, $W$ and $\Delta U$ correspond to those shown in Fig.~\ref{fig: ss and energy}, while the surface area $A$ represents the two-dimensional geometric projection of the crack's surface onto the $x$--$z$ plane. 
Note that the initial drop of $-Q$ in Fig.~\ref{fig: Ediss}(a) \textcolor{black}{(observed at 300\,K, but not at 10\,K)} corresponds to the elastic deformation stage at which the crack surface area remains unchanged at $A=40$\,nm$^{2}$, while the heat is supplied to the system ($Q>0$), as discussed above with reference to Fig.~\ref{fig: ss and energy}.

By following the approach of \citet{XU2020}, the dependence of $-Q$ on $A$ can now be used to determine the critical energy release rate $G_{\rm c}$. Indeed, in the quasi-steady state crack propagation regime, the heat taken from the system by the thermostat corresponds to the energy dissipated as a result of crack propagation and dislocation activity, thus $\mathrm{d}E_{\rm dis}=-\mathrm{d}Q$. For each sample, $G_{\rm c}$ can thus be determined as the slope of the corresponding curve in Fig.~\ref{fig: Ediss} according to Eq.~\eqref{eq:Edis}. Note that the initial stage (marked in blue) and the final stage (marked in pink) are excluded from calculations. The transient initial stage corresponds to the build-up of the plastic zone ahead of the crack tip. The final stage is omitted because of the direct interaction of the crack front with the boundary of the simulation box.

\begin{figure}[t!]
\centering
\includegraphics[scale=0.51]{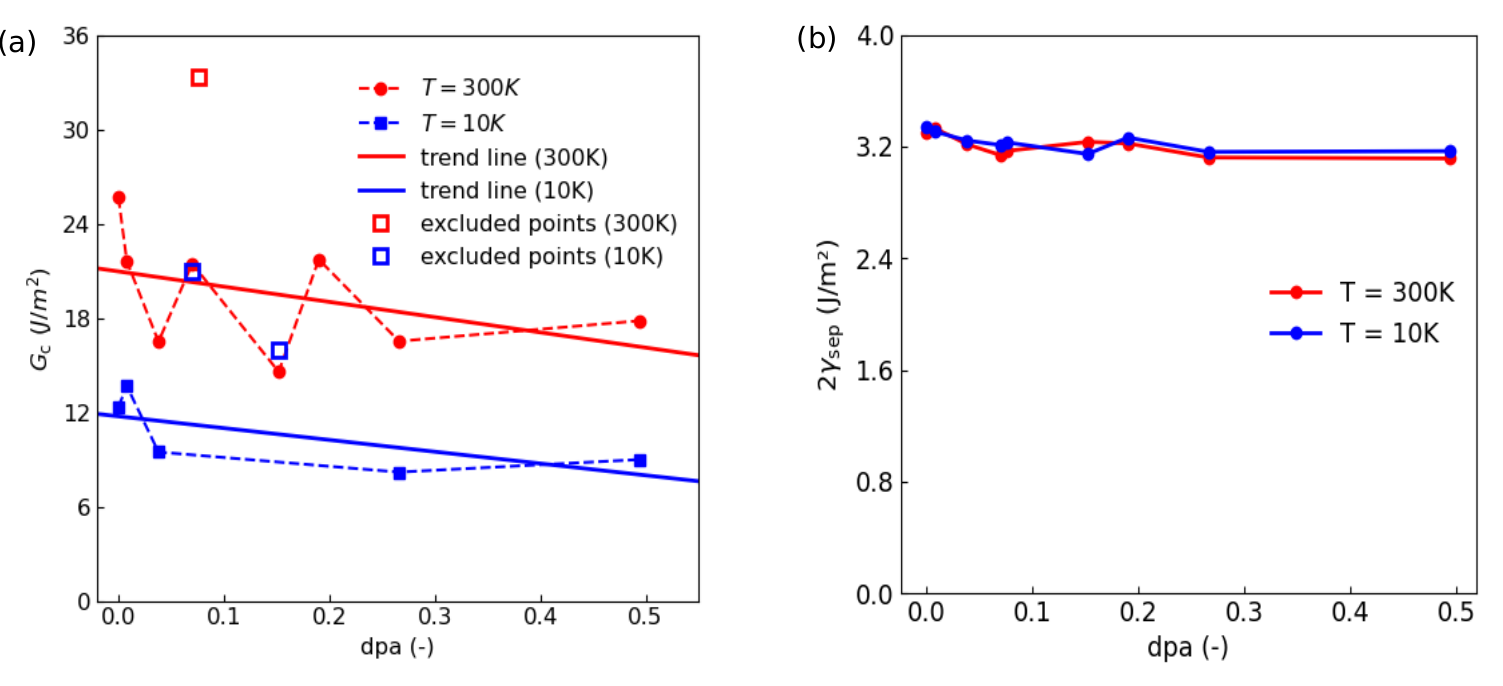}
\caption{(a) Critical energy release rate $G_{\rm c}$ as a function of irradiation level (dpa) at different temperatures: 300K (dashed red line) and 10K (dashed blue line), with solid lines indicating approximation curves. (b) Separation energy 2$\gamma_{\rm sep}$ for crack surface formation as a function of dpa.}
\label{fig: G_c}
\end{figure}

Fig. \ref{fig: G_c}(a) presents the critical energy release rate, $G_{\rm c}$, determined according to Eq. \eqref{eq:Edis}, as a function of irradiation level (dpa) at two different temperatures: $T$=300K (red markers) and $T$=10K (blue markers). The solid lines represent linear approximations of these trends. The extreme values of $G_{\rm c}$ (denoted by empty squares) are excluded to emphasize the dominant irradiation-driven trends. These correspond to cases where the crack encounters strong obstacles, such as entangled dislocations (compare Fig. \ref{fig: blocked crack}), which hinder its propagation and result in anomalously high fracture energy values. Fig. \ref{fig: G_c}(a) shows that $G_{\rm c}$ decreases with increasing dpa, indicating a reduction in the material's resistance to fracture as irradiation damage accumulates. Fig. \ref{fig: G_c}(b) shows the separation energy 2$\gamma_{\rm sep}$ as a function of dpa, defined as the energy required to break atomic bonds and create two fracture surfaces in the absence of plastic deformation. This parameter serves as a reference point for intrinsic surface energy and offers additional insight into the interplay between defect density and fracture energy response. The overall values of 2$\gamma_{\rm sep}$ at 300K and 10K remain comparable and show little variation with dpa, indicating that both temperature and irradiation have a small effect on the intrinsic separation energy in the absence of significant plastic deformation.

\begin{figure}[b!]
\centering
\includegraphics[scale=0.51]{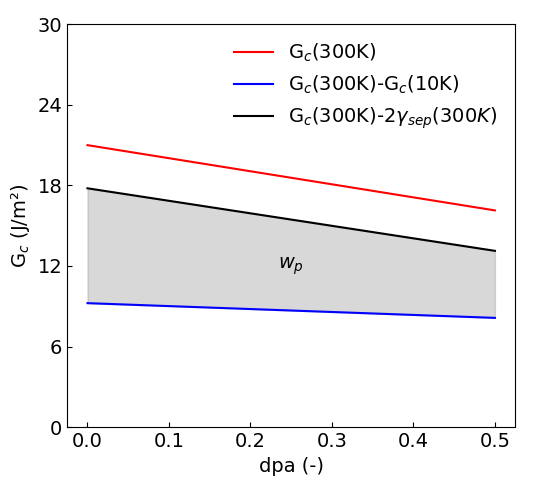}
\caption{Plastic work, $w_{\rm p}$, as a function of the irradiation dose. The red curve represents $G_{\rm c}$(300K), the blue curve shows the difference between the total critical energy release rate at T=300K and the critical energy release rate at T=10K, and the black curve denotes $G_{\rm c}(300 \rm K)-2\gamma_{\rm {sep}}(300\rm K)$, see Fig.~\ref{fig: G_c}. The shaded region highlights the estimated range of values of the plastic work, $w_{\rm p}$.}
\label{fig: w_p}
\end{figure}

The results show a trend consistent with the irradiation-induced DBT. At T=300K, $G_{\rm c}$ decreases over the examined dpa range, indicating suppressed energy dissipation mechanisms as defect density increases. The higher $G_{\rm c}$ at 300K is attributed to thermally activated processes that allow some dislocation-mediated energy absorption. At T=10K, $G_{\rm c}$ moves closer to the theoretical limit governed by intrinsic surface energy $\gamma_{\rm {sep}}$ (Fig. \ref{fig: G_c}(b)), as the lack of thermal energy prevents significant dislocation motion, making bond rupture the predominant fracture mechanism. 

Fig. \ref{fig: w_p} is derived from the data presented in Fig. \ref{fig: G_c} (only linear trend lines are shown) and illustrates the contribution of energy dissipation mechanisms as a function of irradiation level (dpa).
The red curve represents $G_{\rm c}$(300K), which captures the total energy required for crack propagation at 300K. The blue curve, defined as $G_{c}(300 \rm K)-G_{c}(10\rm K)$ represents the lower bound of $w_{\rm p}$, and the black curve defined as $G_{\rm c}(300 \rm K)-2\gamma_{\rm {sep}}(300\rm K)$ represents the upper bound of $w_{\rm p}$. The shaded region highlights the progressive reduction in energy dissipation, $w_{\rm p}$, associated with dislocation-mediated processes as the irradiation dose increases. This reduction indicates the suppression of dislocation activity due to the accumulation of irradiation-induced defects.

\section{Discussion}
\label{sec:conclusion}
\vspace{0.3cm}
This study provides a comprehensive investigation into the fracture behavior of irradiated Fe--Ni--Cr alloys, revealing how radiation-induced defects govern the transition from ductile to brittle failure. 
\textcolor{black}{The simulated defect configuration reflects the damage accumulation characteristic of neutron irradiation, allowing for a direct examination of how radiation-induced defects distributed throughout the material influence crack propagation and reduce the capacity for plastic deformation \citep{AYANOGLU2022, PAREIGE2022}.}
Although the system investigated in this study does not exhibit fully brittle failure due to the limited defect density inherent to the small MD sample size, the observed embrittlement trends offer valuable insights into the underlying mechanisms governing this transition.

The crack propagation behavior in irradiated materials is governed by a complex interplay of several mechanisms, as detailed in Section \ref{sec:33}. Voids act as weak zones that accelerate fracture through coalescence, reducing the cohesive energy and providing pathways for crack growth. Dislocation loops obstruct dislocation motion, redirecting cracks and promoting branching, which alters the crack path and increases fracture surface area. At higher irradiation levels, amorphous zones form at crack tips, suppressing plasticity and promoting a more brittle failure mode. High defect densities also lead to crack branching and the initiation of secondary cracks, which coalesce with the main crack, resulting in irregular fracture patterns. Additionally, radiation-induced defects generate localized stress concentrations, further influencing crack propagation and fracture morphology. These mechanisms collectively contribute to embrittlement, with the balance between plasticity and bond rupture shifting as irradiation damage accumulates.

Atomistic simulations and energy-based analysis demonstrate that irradiation embrittlement arises from the suppression of plasticity, driven by defect-mediated interactions that restrict dislocation motion and alter crack propagation.
To quantify the material’s diminishing capacity for plastic deformation under irradiation, we decompose the critical energy release rate, $G_{\rm c}$, into two primary contributions: surface energy $\gamma_{\rm s}$, which represents the energy required to create new fracture surfaces, and plastic work $w_{\rm p}$, associated with dislocation motion and other inelastic deformations.

We estimate the surface energy $\gamma_{\rm s}$ using two approaches: one derived from the separation energy ($2\gamma_{\rm sep}$) and another from the critical energy release rate at low temperatures ($G_{\rm c}$(10K)), treated as the lower and upper bound of $\gamma_{\rm s}$, respectively. At $T$=10K, where plasticity is suppressed, $G_{\rm c}(10K)$ provides an approximation of the near-brittle fracture limit. To determine the plastic work contribution, we compare the fracture energy at $T$=300K, where plasticity remains active, to the near-brittle reference state at $T$=10K. This allows us to isolate the energy dissipated through plastic deformation as irradiation-induced defects interact with dislocations. Ultimately, the plastic work contribution is specified using two complementary definitions.

This energy-based framework provides a generalizable approach to assessing embrittlement, independent of specific temperature values, as long as the chosen low-temperature condition captures near-brittle behavior while the high-temperature condition allows sufficient plasticity. The results reveal that irradiation progressively reduces plastic work, directly linking irradiation hardening to suppressed dislocation activity. This approach connects microstructural damage to macroscopic fracture resistance, offering a quantitative framework for evaluating fracture behavior in irradiated materials.

\vspace{0.3cm}

The study advances the field through four key contributions:

\begin{itemize}
\item \textit{Defect-driven fracture pathways:~} By categorizing crack--void and crack--loop interactions, we establish that voids accelerate fracture by reducing cohesive strength, while loops impede propagation by pinning dislocations. This duality explains the transition from ductile tearing to brittle cleavage as dpa increases.  
\item \textit{Microstructural--mechanical linkage:~} The correlation between defect density, crack morphology, and energy dissipation bridges atomic-scale simulations to macroscopic fracture resistance, offering a unified model for irradiation embrittlement.
\item \textit{Energy-based embrittlement criterion:~} Building on \citet{XU2020}, we adapt the decomposition of $G_{\rm c}$ into brittle ($\gamma_{\rm s}$) and plastic ($w_{\rm p}$) contributions for irradiated materials. This framework provides a step towards quantifying how irradiation affects the fracture energy balance.
\item \textit{Quantitative description of irradiation-induced DBT:~} The study demonstrates a decrease in both $G_{\rm c}$ and $w_{\rm p}$ with increasing dpa, providing a quantitative manifestation of the irradiation-induced DBT. While the observed embrittlement is relatively small due to the limited dpa range in our MD simulations, real-world scenarios with higher dpa values and greater defect densities are expected to exhibit more pronounced DBT effects.
\end{itemize}

\textcolor{black}{These findings pertain specifically to austenitic Fe--Ni--Cr alloys with an fcc lattice, where defect distributions are largely isotropic due to uniform threshold displacement energies and minimal channeling effects. Ferritic-martensitic steels, also used in nuclear fusion applications, are characterized by a bcc structure and low nickel content to mitigate transmutation reactions and reduce helium accumulation. These differences may influence the overall crack behavior because the variations in microstructure and defect evolution may modulate the fracture resistance. Nonetheless, the core mechanisms driving irradiation-induced embrittlement, dislocation pinning, void coalescence, and crack branching are expected to be fundamentally similar in both fcc and bcc steels. Thus, while this work directly addresses fcc austenitic steels, its insights contribute to a broader understanding of irradiation embrittlement.}

\section{Conclusions}
\label{sec:conclusion}
\vspace{0.3cm}
\textcolor{black}{This work has systematically investigated the fracture mechanisms in irradiated alloys, emphasizing the crucial role of radiation-induced defects in embrittlement. 
Based on MD simulations, the analysis shows that irradiation decreases the material’s ability to accommodate plastic deformation by impeding dislocation motion and promoting crack growth through the absorption of radiation-induced voids. These voids act as weak zones that facilitate crack propagation, collectively causing a transition from ductile to more brittle fracture behavior.}

\textcolor{black}{Our energy decomposition framework determines embrittlement by correlating atomistic defect-induced changes in fracture energy components with the macroscopic loss of plastic dissipation capacity. By separating total fracture energy into surface energy (crack surface creation) and plastic work (dislocation-mediated inelastic deformation), it provides a robust parameter that translates atomic-scale damage effects into continuum-scale fracture behavior.}


\section*{Acknowledgments}
The National Science Centre in Poland supports this work through Grant No UMO-2020/38/E/ST8/00453. 
FJDG  acknowledges partial support from the European Union Horizon 2020 research and innovation program under grant agreement No. 857470 and from the European Regional Development Fund via the Foundation for Polish Science International Research Agenda PLUS program grant No. MAB PLUS/2018/8. We gratefully acknowledge Polish high-performance computing infrastructure PLGrid (HPC Center at ACK Cyfronet AGH) for providing computer facilities and support within computational Grant No. PLG/2024/017084 and the
CIS High Performance Cluster at the National Centre for Nuclear Research in Poland.


\appendix
\renewcommand{\thefigure}{A.\arabic{figure}}
\setcounter{figure}{0}
\section{Evolution of defect densities in irradiated samples}
\label{sec:appendixA}
\textcolor{black}{Supplementary data provided in this appendix illustrate the evolution of radiation-induced defect densities in the irradiated samples discussed in Section \ref{sec: Section2.1}.}

\begin{figure}[b!]
\centering
\includegraphics[width=0.43\textwidth]{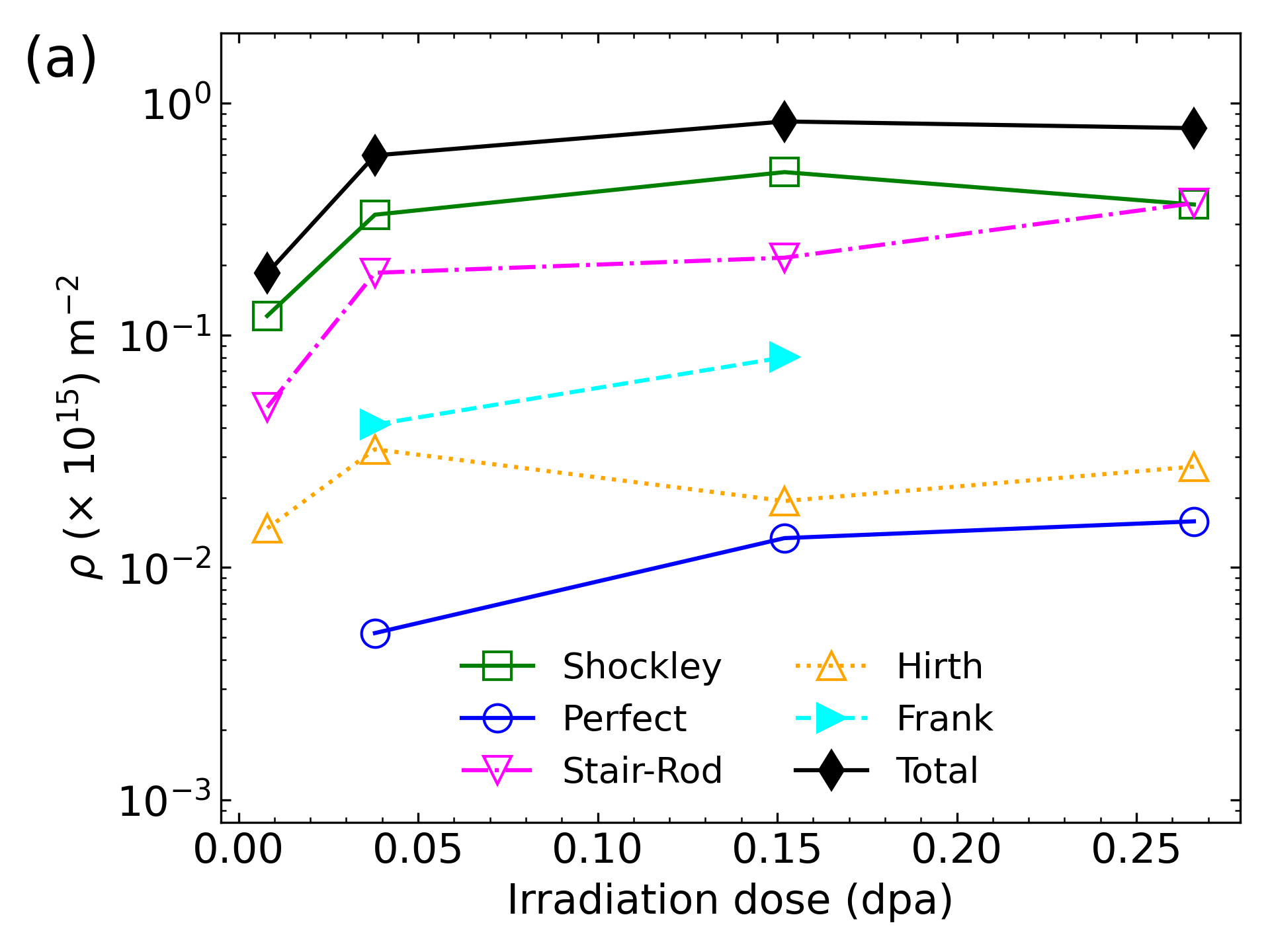}
\includegraphics[width=0.43\textwidth]{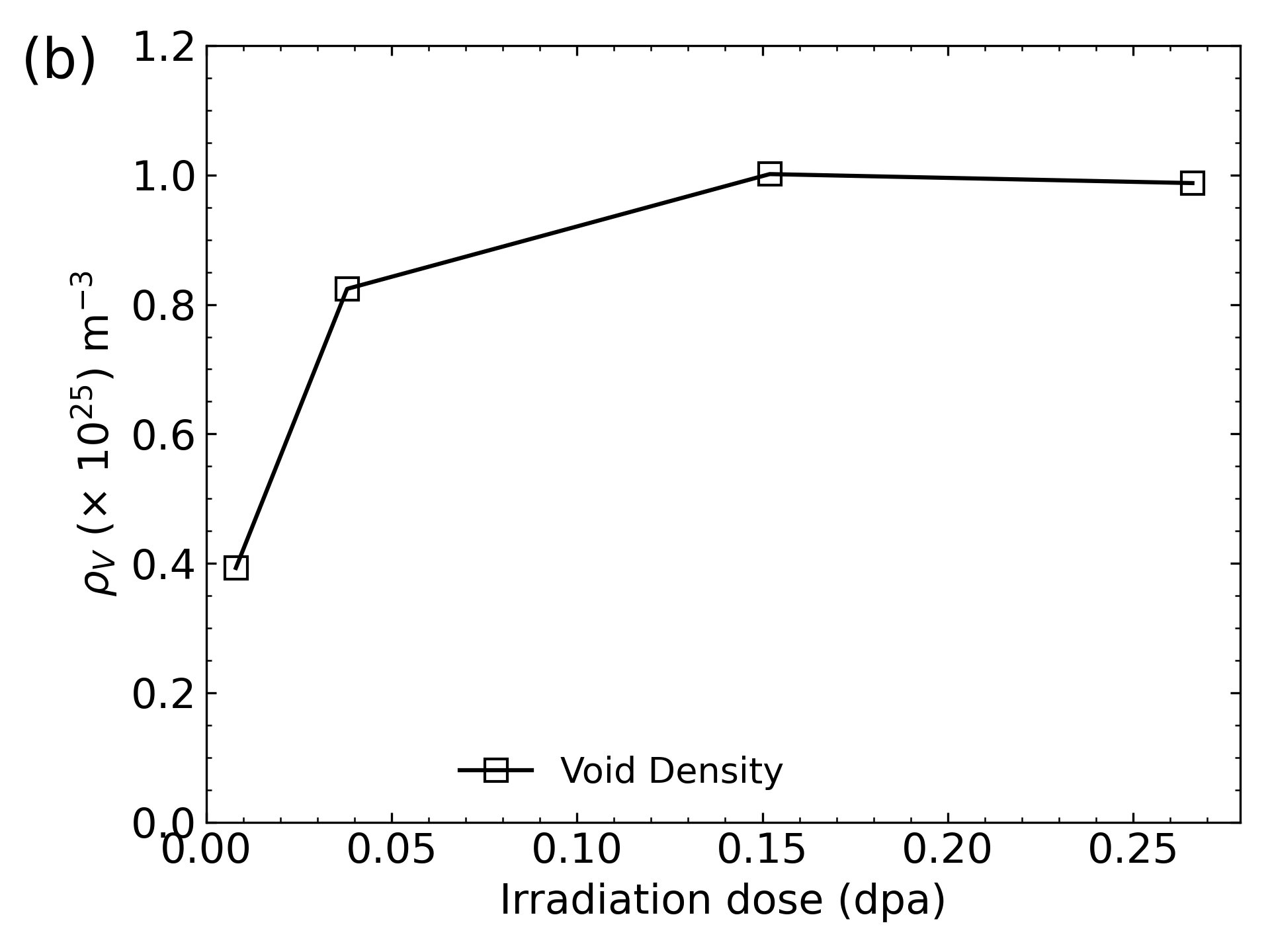}
    \caption{Density of dislocation loops (a) and voids (b) as a function 
    of the irradiation dose in dpa obtained from MD simulations of collision cascades.}
    \label{fig:defects density}
\end{figure}

\textcolor{black}{Figure~\ref{fig:defects density} shows how the densities of radiation-induced dislocation loops and voids evolve with increasing irradiation dose. The density of radiation defects increases gradually at lower dpa, reflecting the early stages of dislocation loop nucleation and the initial aggregation of vacancies into stable void structures. \textcolor{black}{The absence of perfect dislocations at dpa = 0.008 reflects the early stage of defect accumulation, where primarily Shockley partials and small defect clusters are present. Perfect dislocations tend to form at higher dpa levels as overlapping cascades drive defect evolution. Void density includes all voids regardless of size.} At higher dpa levels, the accumulation and interaction of dislocations lead to the formation of various dislocation types and a more complex defect network.}

\renewcommand{\thefigure}{B.\arabic{figure}}
\setcounter{figure}{0}
\section{Effect of boundary conditions on fracture behavior}
\label{sec:appendixB}

\begin{figure}[b!]
    \centering
    \includegraphics[width=\textwidth]{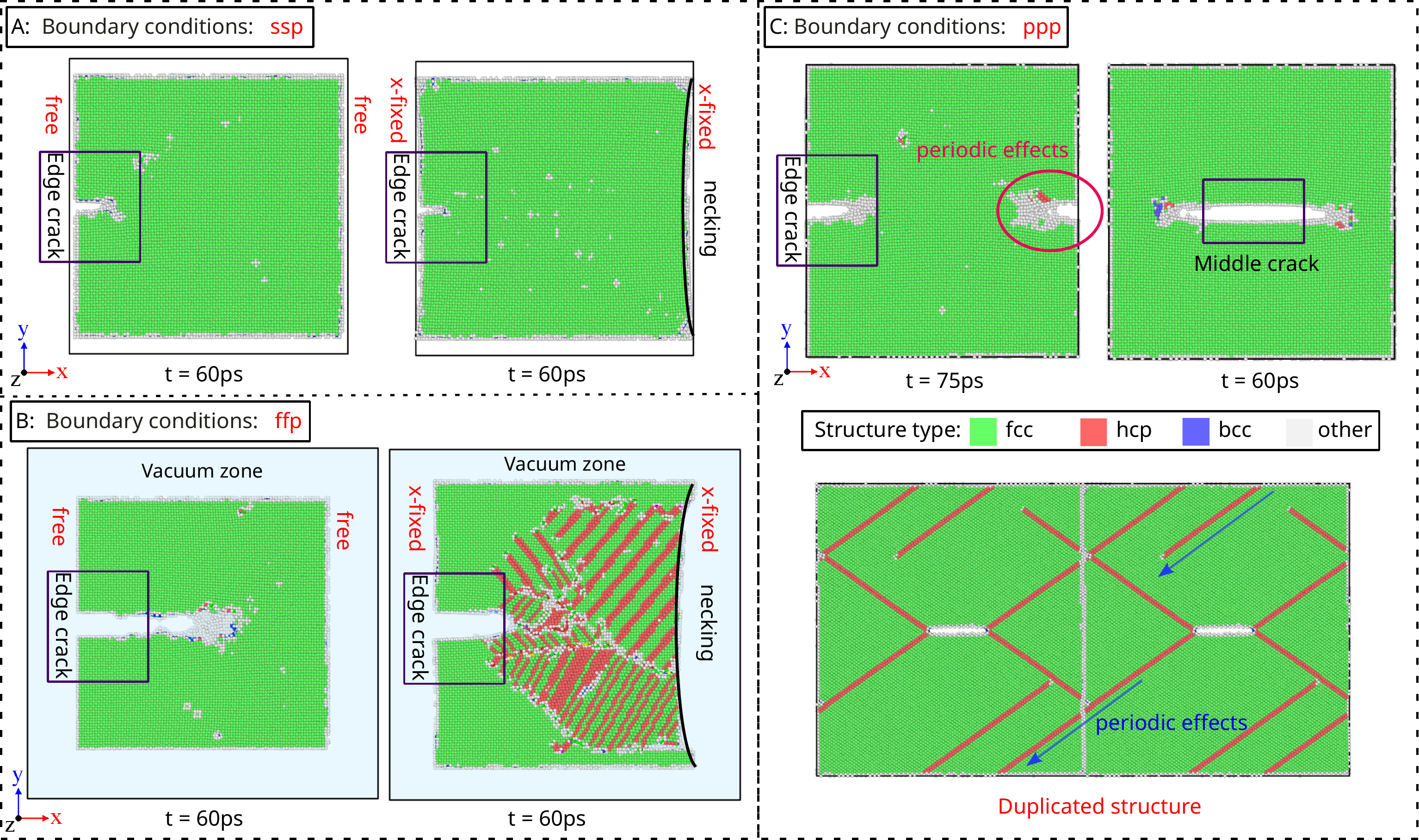}
\caption{\textcolor{black}{Effect of boundary conditions on crack propagation. 
Top-left panel A show two variations of the \texttt{ssp} configuration: 
(i) with fixed lateral boundaries (left), corresponding to the setup used in the main study, and 
(ii) with free lateral boundaries (right). Bottom-left panel B shows the \texttt{ffp} configuration: (i) with constrained lateral edges and a vacuum region (light blue), and (ii) with unconstrained sides leading to pronounced necking. Panel C on the right shows the \texttt{ppp} setup with periodic boundary conditions for two crack locations: at the sample edge (left) and at the center (right). These conditions lead to periodic effects such as mirrored crack propagation (purple) and structural duplication (blue).}}
\label{fig:boundary_conditions}
\end{figure}
\textcolor{black}{To assess how boundary constraints affect the physical accuracy of crack propagation in atomistic simulations, we performed a comparative analysis of three representative boundary setups commonly used in MD studies. Specifically, we examined: (i) shrink-wrapped lateral boundaries with fixed displacements applied at the top and bottom (ssp), (ii) fixed boundaries in all directions with vacuum padding (ffp), and (iii) fully periodic boundary conditions (ppp). All simulations were conducted under uniaxial tensile loading. Representative snapshots of the resulting deformation patterns and fracture evolution are provided in Fig.~\ref{fig:boundary_conditions}. Such boundary configurations have also been investigated in previous studies focusing on fracture behavior, including fully periodic (ppp) \citep{SINGH2019}, fully free (ffp), and non-periodic shrink-wrapped (ssp) systems \citep{Krull2011, XU2020, DING2020, BARIK2023}.}

\textcolor{black}{The upper-left panel A of Fig.~\ref{fig:boundary_conditions} illustrates two variations of the ssp configuration. The left snapshot in panel A corresponds to the ssp setup employed throughout our study, where atoms at the lateral edges (in the $x$-direction) are constrained to prevent transverse motion. In contrast, the right snapshot in panel~A depicts a similar ssp configuration without lateral constraints, resulting in pronounced necking near the edges. The released lateral strain causes localized narrowing, which impedes the crack front and obstructs its advancement.}

\textcolor{black}{The ssp setup imitates a tensile test with clamped top and bottom edges, closely resembling experimental conditions where deformation is restricted by grips. The applied displacement is transmitted through fixed atomic layers at the top and bottom of the sample, inducing a deformation that gradually propagates through the lattice.} 

\textcolor{black}{In contrast, the ffp setup (lower-left panel B) enforces affine deformation by globally modifying the simulation box dimensions. While this method is efficient and stable, it imposes a homogeneous strain field that accelerates local crack opening. The applied global strain leads to earlier failure initiation and can obscure important mechanisms such as strain localization or dislocation emission. This approach can approximate bulk material behavior, and the imposed uniform deformation limits local surface-related effects.}

\textcolor{black}{The lower-left panel B of Fig.~\ref{fig:boundary_conditions} illustrates two 
variants of the ffp boundary condition. The left snapshot in panel B corresponds to the ffp setup where atoms at the lateral edges (in the $x$-direction) are constrained to prevent transverse motion. To prevent atom deletion at the sample edges, a vacuum region (light blue) is introduced. This setup enforces affine deformation by globally modifying the simulation box dimensions. While efficient and stable, it imposes a homogeneous strain field that accelerates local crack opening and leads to necking near the constrained edges (compare snapshots for the same time steps in panels A and~B). The applied global strain causes earlier failure initiation and can obscure important mechanisms such as strain localization or dislocation emission. In contrast, the right snapshot in panel~B shows a similar ffp configuration without lateral constraints, leading to more freedom at the edges and resulting in more pronounced necking and different deformation behavior. This approach can approximate bulk material behavior. The imposed uniform deformation limits local surface-related effects and does not fully replicate free surface conditions essential for fracture studies.}

\textcolor{black}{The large panel on the right side of Fig.~\ref{fig:boundary_conditions} corresponds to the \texttt{ppp} configuration. Two cases are shown: with the crack initially placed at the left edge (top left) and at the center of the sample (top right). Due to periodic boundary conditions, mirrored cracks appear on the opposite side (highlighted in purple), and structural features are periodically duplicated (marked in blue). These effects modify the deformation characteristics and should be considered in the overall assessment of the system’s behavior.}

\textcolor{black}{The constrained boundary ssp setup used in this work has been selected as a compromise that preserves realistic plasticity and dislocation behavior while preventing unphysical deformation modes.}

\renewcommand{\thefigure}{C.\arabic{figure}}
\setcounter{figure}{0}
\section{Strain field maps of irradiated and unirradiated samples}
\label{sec:appendixC}
\begin{figure}[t!]
\centering
\includegraphics[scale=0.4]{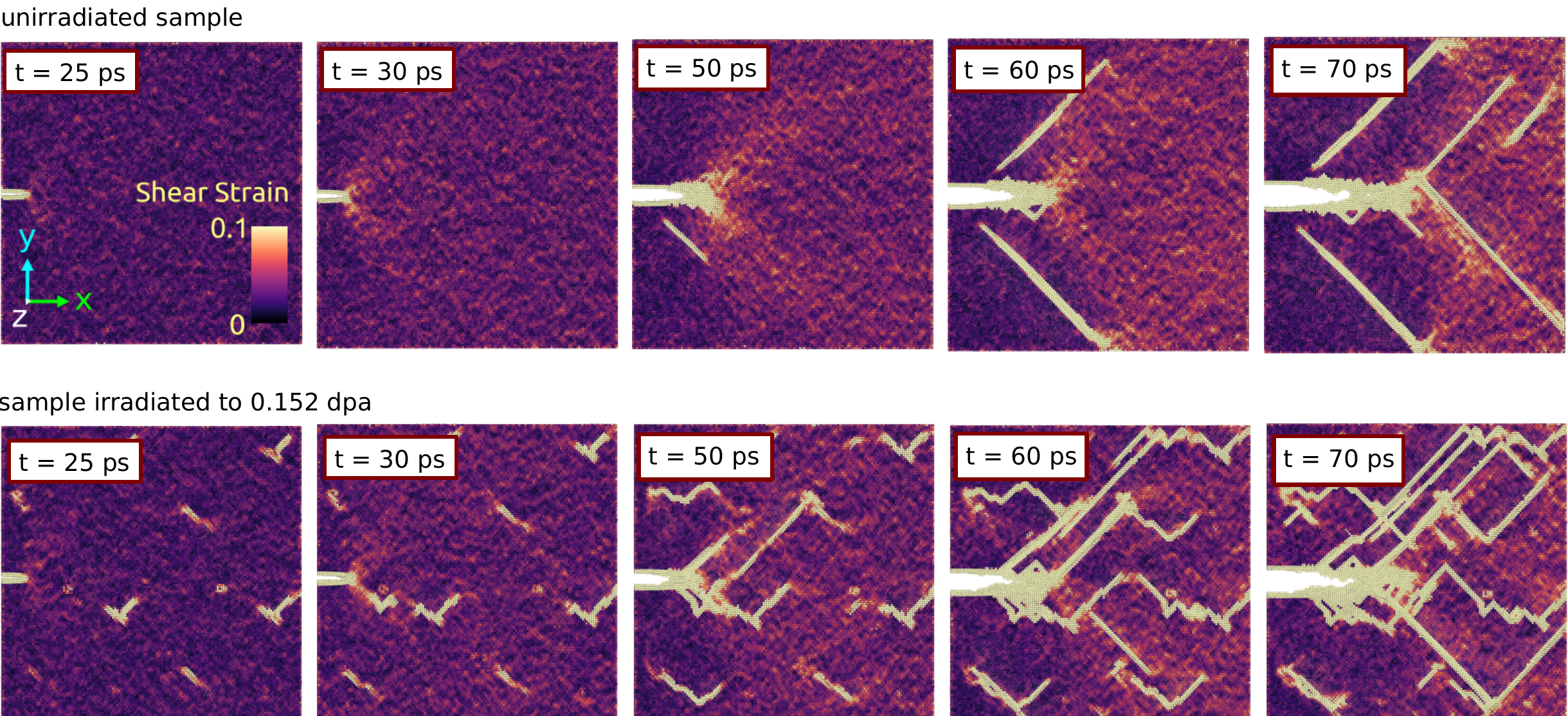}
\caption{\textcolor{black}{Evolution of the equivalent shear strain (von Mises invariant of the Green--Lagrangian strain tensor~\citep{Stukowski_2010})} in a selected cross-section of unirradiated (top) and irradiated (bottom) samples.}
 \label{fig: strain fields}
\end{figure}
\textcolor{black}{This appendix presents supplementary strain field maps used to visualize deformation behavior in the irradiated and unirradiated samples discussed in Section \ref{sec: Section2.2}.
Fig.~\ref{fig: strain fields} presents equivalent shear strain fields in both unirradiated and irradiated samples across various time steps. The equivalent shear strain is calculated by OVITO as the von Mises invariant of the Green--Lagrangian strain tensor~\citep{Stukowski_2010}. The images show clear differences in strain localization patterns, where the irradiated samples exhibit concentrated strain fields at radiation-induced defects.}

\bibliographystyle{unsrtnat}
\renewcommand*{\bibfont}{\fontsize{10pt}{10pt}}
\bibliography{references}

\end{document}